\documentclass[aps,prd,a4paper,twocolumn,amsmath,showpacs,superscriptaddress,nofootinbib,preprintnumbers]{revtex4-1}
\usepackage{graphicx,ulem}
\usepackage{longtable}
\usepackage{float}
\usepackage{dcolumn}
\usepackage{graphics,epsfig}
\usepackage{amsmath,amssymb,latexsym,mathrsfs}
\usepackage{bm}
\usepackage{color}
\usepackage{color}
\usepackage{subfigure}
\usepackage{hyperref}
\usepackage{wasysym}\usepackage{array}
\newcolumntype{?}{!{\vrule width 5pt}}

\newcommand{\eV}{\,\mathrm{eV}}

\newcommand{\Mnumin}{M_{\nu,\mathrm{min}}}
\usepackage{titlesec}
\titleformat{\paragraph}
{\normalfont\normalsize\bfseries}{\theparagraph}{1em}{}
\titlespacing*{\paragraph}
{0pt}{3.25ex plus 1ex minus .2ex}{1.5ex plus .2ex}

\begin{document}

\title{Unveiling $\nu$ secrets with cosmological data: neutrino masses and mass hierarchy}

\author{Sunny Vagnozzi}
\email{sunny.vagnozzi@fysik.su.se}
\affiliation{The Oskar Klein Centre for Cosmoparticle Physics, Department of Physics, Stockholm University, SE-106 91 Stockholm, Sweden}

\author{Elena Giusarma}
\email{egiusarm@andrew.cmu.edu}
\affiliation{McWilliams Center for Cosmology, Department of Physics, Carnegie Mellon University, Pittsburgh, PA 15213, USA}
\affiliation{Lawrence Berkeley National Laboratory (LBNL), Physics Division, Berkeley, CA 94720-8153, USA}
\affiliation{Berkeley Center for Cosmological Physics, University of California, Berkeley, CA 94720, USA}

\author{Olga Mena}
\email{omena@ific.uv.es}
\affiliation{Instituto de F\'{i}sica Corpuscolar (IFIC), Universidad de Valencia-CSIC, E-46980, Valencia, Spain}

\author{Katherine Freese}
\affiliation{The Oskar Klein Centre for Cosmoparticle Physics, Department of Physics, Stockholm University, SE-106 91 Stockholm, Sweden}
\affiliation{Michigan Center for Theoretical Physics, Department of Physics, University of Michigan, Ann Arbor, MI 48109, USA}

\author{Martina Gerbino}
\affiliation{The Oskar Klein Centre for Cosmoparticle Physics, Department of Physics, Stockholm University, SE-106 91 Stockholm, Sweden}

\author{Shirley Ho}
\affiliation{McWilliams Center for Cosmology, Department of Physics, Carnegie Mellon University, Pittsburgh, PA 15213, USA}
\affiliation{Lawrence Berkeley National Laboratory (LBNL), Physics Division, Berkeley, CA 94720-8153, USA}
\affiliation{Berkeley Center for Cosmological Physics, University of California, Berkeley, CA 94720, USA}

\author{Massimiliano Lattanzi}
\affiliation{Dipartimento di Fisica e Scienze della Terra, Universit\`a di Ferrara, I-44122 Ferrara, Italy}
\affiliation{Istituto Nazionale di Fisica Nucleare (INFN), Sezione di Ferrara, I-44122 Ferrara, Italy}

\date{\today}

\begin{abstract}
Using some of the latest cosmological datasets publicly available, we derive the strongest bounds in the literature on the sum of the three active neutrino masses, $M_\nu$, within the assumption of a background flat $\Lambda$CDM cosmology. In the most conservative scheme, combining Planck cosmic microwave background (CMB) temperature anisotropies and baryon acoustic oscillations (BAO) data, as well as the up-to-date constraint on the optical depth to reionization ($\tau$), the tightest $95\%$ confidence level (C.L.) upper bound we find is $M_\nu<0.151$~eV. The addition of Planck high-$\ell$ polarization data, which however might still be contaminated by systematics, further tightens the bound to $M_\nu<0.118$~eV. A proper model comparison treatment shows that the two aforementioned combinations disfavor the IH at $\sim 64\%$~C.L. and $\sim 71\%$~C.L. respectively.  In addition, we compare the constraining power of measurements of the full-shape galaxy power spectrum versus the BAO signature, from the BOSS survey. Even though the latest BOSS full shape measurements cover a larger volume and benefit from smaller error bars compared to previous similar measurements, the analysis method commonly adopted results in their constraining power still being less powerful than that of the extracted BAO signal. Our work uses only cosmological data; imposing the constraint $M_\nu>0.06\,{\rm eV}$ from oscillations data would raise the quoted upper bounds by ${\cal O}(0.1\sigma)$ and would not affect our conclusions.
\end{abstract}

\maketitle

\section{Introduction}\label{sec:intro}

The discovery of neutrino oscillations, which resulted in the 2015 Nobel Prize in Physics~\cite{nobel}, has robustly established the fact that neutrinos are massive~\cite{superk,sno,kamland,minos,dayabay,reno,doublechooz,t2k}. The results from oscillation experiments can therefore be successfully explained assuming that the three neutrino flavour eigenstates ($\nu_e$, $\nu_{\mu}$, $\nu_{\tau}$) are quantum superpositions of three mass eigenstates ($\nu_1$, $\nu_2$, $\nu_3$). In analogy to the quark sector, flavour and mass eigenstates are related via a mixing matrix parametrized by three mixing angles ($\theta_{12}$, $\theta_{13}$, $\theta_{23}$) and a CP-violating phase $\delta_{\text{CP}}$.

Global fits~\cite{fit1,fit2,fit3,fit4,fit5} to oscillation measurements have determined with unprecedented accuracy five mixing parameters, namely, $\sin^2\theta_{12}$, $\sin^2\theta_{13}$, $\sin^2\theta_{23}$, as well as the two mass-squared splittings governing the solar and the atmospheric transitions. The solar mass-squared splitting is given by $\Delta m_{21}^2 \equiv m^2_2 - m^2_1 \simeq 7.6\times 10^{-5}\,\eV^2$. Because of matter effects in the Sun, we know that the mass eigenstate with the larger electron neutrino fraction is the one with the smallest mass. We identify the lighter state with ``$1$" and the heavier state (which has a smaller electron neutrino fraction) with ``$2$". Consequently, the solar mass-squared splitting is positive. The atmospheric mass-squared splitting is instead given by $|\Delta m_{31}^2| \equiv |m^2_3 - m^2_1|\simeq 2.5\times 10^{-3}\eV^2$. Since the sign of the largest mass-squared splitting $|\Delta m_{31}^2|$ remains unknown, there are two possibilities for the mass ordering: the normal hierarchy (NH, $\Delta m_{31}^2 > 0$, with $m_1<m_2<m_3$) and the inverted hierarchy (IH, $\Delta m_{31}^2 < 0$, and $m_3<m_1<m_2$). Other unknowns in the neutrino sector are the presence of CP-violation effects (i.e. the value of $\delta_{\text{CP}}$), the $\theta_{23}$ octant, the Dirac versus Majorana neutrino nature, and, finally, the absolute neutrino mass scale, see Ref.~\cite{Capozzi:2016rtj} for a recent review on unknowns of the neutrino sector.

Cosmology can address two out of the above five unknowns: the absolute mass scale and the mass ordering. Through background effects, cosmology is to zeroth-order sensitive to the absolute neutrino mass scale, that is, to the quantity:
\begin{eqnarray}
M_\nu \equiv m_{\nu_1} + m_{\nu_2} + m_{\nu_3} \, ,
\end{eqnarray}
where $m_{\nu_i}$ denotes the mass of the $i$th neutrino mass eigenstate. Indeed, the tightest current bounds on the neutrino mass scale come from cosmological probes, see for instance \cite{Giusarma:2014zza,Palanque-Delabrouille:2015pga,DiValentino:2015wba,DiValentino:2015sam,Cuesta:2015iho,Huang:2015wrx,DiValentino:2016ikp,Giusarma:2016phn}. More subtle perturbation effects make cosmology in principle sensitive to the mass hierarchy as well (see e.g. \cite{Lesgourgues:2006nd,Wong:2011ip,Lesgourgues:2012uu,Abazajian:2013oma,book,Lesgourgues:2014zoa} for comprehensive reviews on the impact of nonzero neutrino masses on cosmology), although not with current datasets.

As light massive particles, relic neutrinos are relativistic in the early Universe and contribute to the radiation energy density. However, when they turn non-relativistic at late times, their energy density contributes to the total matter density. Thus, relic neutrinos leave a characteristic imprint on cosmological observables, altering both the background evolution and the spectra of matter perturbations and Cosmic Microwave Background (CMB) anisotropies (see \cite{Lesgourgues:2006nd,Wong:2011ip,Lesgourgues:2012uu,Abazajian:2013oma,book,Lesgourgues:2014zoa} as well as the recent \cite{Archidiacono:2016lnv} for a detailed review on massive neutrinos in cosmology, in light of both current and future datasets). The effects of massive neutrinos on cosmological observables will be discussed in detail in Sec.~\ref{sec:datasets}.

Cosmological probes are primarily sensitive to the sum of the three active neutrino masses $M_\nu$. The exact distribution of the total mass among the three mass eigenstates induces sub-percent effects on the different cosmological observables, which are below the sensitivities of ongoing and near future experiments~\cite{Lesgourgues:2004ps,Pritchard:2008wy,DeBernardis:2009di,Jimenez:2010ev,Wagner:2012sw}. As a result, cosmological constraints on $M_\nu$ are usually obtained by making the assumption of a fully degenerate mass spectrum, with the three neutrinos sharing the total mass [$m_{\nu_i}=M_\nu/3$, with $i=1,2,3$, which we will later refer to as \textit{3deg}, see Eq.(\ref{distributions})]. Strictly speaking, this is a valid approximation as long as the mass of the lightest eigenstate, $m_0 \equiv m_1 \ [m_3]$ in the case of NH [IH], satisfies:
\begin{eqnarray}
m_0 \gg \vert m_i-m_j \vert \quad \, , \quad \forall i,j=1,2,3.
\end{eqnarray}
The approximation might fail in capturing the exact behaviour of massive neutrinos when $M_\nu\sim\Mnumin$, where $\Mnumin=\sqrt{\Delta m_{21}^2}+\sqrt{\Delta m_{31}^2}\simeq0.06\,\eV\,[=\sqrt{\Delta m_{31}^2}+\sqrt{\Delta m_{31}^2+\Delta m_{21}^2}\simeq0.1\,\eV]$ is the minimal mass allowed by oscillation measurements in the NH [IH] scenario~\cite{fit1,fit2,fit3,fit4,fit5}, see Appendix A for detailed discussions. Furthermore, it has been argued that the ability to reach a robust upper bound on the total neutrino mass below $\Mnumin=0.1\,\eV$ would imply having discarded at some statistical significance the inverted hierarchy scenario. In this case, one has to provide a rigorous statistical treatment of the preference for one hierarchy over the other~\cite{Hannestad:2016fog,Xu:2016ddc,martinanew}. \textbf{3deg repeated}

We will be presenting results obtained within the approximation of three massive degenerate neutrinos. That is, we consider the following mass scheme, which we refer to as \textit{3deg}:
\begin{eqnarray}
m_1 = m_2 = m_3 = \frac{M_\nu}{3} \quad (\textbf{\textit{3deg}}) \, , \nonumber
\label{distributions}
\end{eqnarray}
This approximation has been adopted by the vast majority of works when $M_\nu$ is allowed to vary. This includes the Planck collaboration, which recently obtained $M_{\nu} < 0.234 \ {\rm eV}$ at 95\%~C.L. \cite{planckcosmological} through a combination of temperature and low-$\ell$ polarization anisotropy measurements, within the assumption of a flat $\Lambda$CDM+$M_\nu$ cosmology. Physically speaking, this choice is dictated by the observation that the impact of the NH and IH mass splittings on cosmological data is tiny if one compares the \textit{3deg} approximation to the corresponding NH and IH models with the same value of the total mass $M_\nu$ (see Appendix A for further discussions). For the purpose of comparison with previous work, in Appendix B we briefly discuss other less physical approximations which have been introduced in the recent literature, as well as some of the bounds obtained on $M_{\nu}$ within such approximations.

We present the constraints in light of the most recent cosmological data publicly available. In particular, we make use of i) measurements of the temperature and polarization anisotropies of the CMB as reported by the Planck satellite in the 2015 data release; ii) baryon acoustic oscillations (BAO) measurements from the SDSS-III BOSS data release 11 CMASS and LOWZ samples, and from the Six-degree Field Galaxy Survey (6dFGS) and WiggleZ surveys; iii) measurements of the galaxy power spectrum of the CMASS sample from the SDSS-III BOSS data release 12; iv) local measurements of the Hubble parameter ($H_0$) from the Hubble Space Telescope; v) the latest measurement of the optical depth to reionization ($\tau$) coming from the analysis of the high-frequency channels of the Planck satellite, and vi) cluster counts from the observation of the thermal Sunyaev-Zeldovich (SZ) effect by the Planck satellite.

In addition to providing bounds on $M_\nu$, we also use these bounds to provide a rigorous statistical treatment of the preference for the NH over the IH. We do so by applying the simple but rigorous method proposed in \cite{Hannestad:2016fog}, and evaluate both posterior odds for NH against IH, as well as the C.L. at which current datasets can disfavor the IH.

The paper is organized as follows. In Sec.~\ref{sec:method}, we describe our analysis methodology. In Sec.~\ref{sec:datasets} we instead provide a careful description of the datasets employed, complemented with a full explanation of the physical effects of massive neutrinos on each of them. We showcase our main results in Sec.~\ref{sec:mnu}, with Sec.~\ref{sec:geometricalandshape} in particular devoted to an analysis of the relative constraining power of shape power spectrum versus geometrical BAO measurements, whereas in Sec.~\ref{sec:modelcomparison} we provide a rigorous quantification of the exclusion limits on the inverted hierarchy from current datasets. Finally, we draw our conclusions in Sec. \ref{sec:conclusion}. 

For the reader who wants to skip to the results: the most important results of this paper can be found in Tabs.~\ref{tab:tabmnubao},~\ref{tab:tabmnupolbao},~\ref{tab:inverted}. The first two of these tables present the most constraining 95\% C.L. bounds on the sum of the neutrino masses using a combination of CMB (temperature and polarization), BAO, and other external datasets. The bounds in Tab.~\ref{tab:tabmnupolbao} have been obtained using also small-scale CMB polarization data which may be contaminated by systematics, yet we present the results as they are useful for comparing to previous work. Finally Tab.~\ref{tab:inverted} presents exclusion limits on the Inverted Hierarchy neutrino mass ordering, which is disfavored at about 70\%~C.L. statistical significance.

\section{Analysis method}
\label{sec:method}

In the following we shall provide a careful description of the statistical methods employed in order to obtain the bounds on the sum of the three active neutrino masses we show in Sec.~\ref{sec:mnu}, as well as caveats to our analyses. Furthermore, we provide a brief description of the statistical method adopted to quantify the exclusion limits on the IH from our bounds on $M_\nu$. For more details on the latter, we refer the reader to~\cite{Hannestad:2016fog} where this method was originally described.

\subsection{Bounds on the total neutrino mass}
\label{sec:bounds}

In our work, we perform standard Bayesian inference (see e.g. \cite{bayes,trottareview} for recent reviews) to derive constraints on the sum of the three active neutrino masses. That is, given a model described by the parameter vector $\boldsymbol{\theta}$, and a set of data $\boldsymbol{x}$, we derive posterior probabilities of the parameters given the data, $p(\boldsymbol{\theta} \vert \boldsymbol{x})$, according to:
\begin{eqnarray}
p(\boldsymbol{\theta} \vert \boldsymbol{x}) \propto {\cal L}(\boldsymbol{x} \vert \boldsymbol{\theta})p(\boldsymbol{\theta}) \, ,
\end{eqnarray}
where ${\cal L}(\boldsymbol{x} \vert \boldsymbol{\theta})$ is the likelihood function of the data given the model parameters, and $p(\boldsymbol{\theta})$ denotes the data-independent prior. We derive the posteriors using the Markov Chain Monte Carlo (MCMC) sampler \texttt{cosmomc} with an efficient sampling method~\cite{cosmomc1,cosmomc2}. To assess the convergence of the generated chains, we employ the Gelman and Rubin statistics \cite{gelmanandrubin} $R-1$, which we require to satisfy $R-1<0.01$ when the datasets do not include SZ cluster counts, $R-1<0.03$ otherwise (this choice is dictated by time and resource considerations: runs involving SZ cluster counts are more computationally expensive than those that do not include SZ clusters, to achieve the same convergence). In this way, the contribution from statistical fluctuations is roughly a few percent the limits quoted.~\footnote{Notice that this is a very conservative requirement, as a convergence of $0.05$ is typically more than sufficient for the exploration of the posterior of a parameter whose distribution is unimodal~\cite{lewiscosmocoffee}.}

We work under the assumption of a background flat $\Lambda$CDM Universe, and thus consider the following seven-dimensional parameter vector:
\begin{eqnarray}
\boldsymbol{\theta} \equiv \{\Omega_b h^2, \Omega _c h^2, \Theta_s, \tau, n_s, \log(10^{10}A_s), M_\nu \} \, .
\label{parameters}
\end{eqnarray}
Here, $\Omega_bh^2$ and $\Omega_ch^2$ denote the physical baryon and dark matter energy densities respectively, $\Theta_s$ is the ratio of the sound horizon to the angular diameter distance at decoupling, $\tau$ indicates the optical depth to reionization, whereas the details of the primordial density fluctuations are encoded in the amplitude ($A_s$) and the spectral index ($n_s$) of its power spectrum at the pivot scale $k_{\star} = 0.05 \ h \ {\rm Mpc} ^{-1}$. Finally, the sum of the three neutrino masses is denoted by $M_\nu$. For all these parameters, a uniform prior is assumed unless otherwise specified.

Concerning $M_\nu$, we impose the requirement $M_\nu\ge 0$. Thus, we ignore prior information from oscillation experiments, which, as previously stated, set a lower limit of $\Mnumin\sim0.06\,\eV\,[0.10\,\eV]$ for the NH [IH] mass ordering. If we instead had chosen not to ignore prior information from oscillation experiments, the result would be a slight shift of the center of mass of our posteriors on $M_\nu$ towards higher values. As a consequence of these shifts, the 95\%~C.L. upper limits we report would also be shifted to slightly higher values. Nonetheless, in this way we can obtain an independent upper limit on $M_\nu$ from cosmology alone, while at the same time making the least amount of assumptions. It also allows us to remain open to the possibility of cosmological models predicting a vanishing neutrino density today, or models where the effect of neutrino masses on cosmological observables is hidden due to degeneracies with other parameters (see e.g.~\cite{Beacom:2004yd,bellini}). One can get a feeling for the size of the shifts by comparing our results to those of \cite{DiValentino:2015sam}, where a prior $M_\nu \geq 0.06\,\eV$ was assumed. As we see, the size of the shifts is small, of ${\cal O}(0.1\sigma)$. We summarize the priors on cosmological parameters, as well as some of the main nuisance parameters, in Tab.~\ref{tab:priors}.

All the bounds on $M_\nu$ reported in Sec.~\ref{sec:mnu} are 95\%~C.L. upper limits. These bounds depend more or less strongly on our assumption of a background flat $\Lambda$CDM model, and would differ if one were to consider extended parameter models, for instance scenarios in which the number of relativistic degrees of freedom $N_{\text{eff}}$ and/or the dark energy equation of state $w$ are allowed to vary, or if the assumption of flatness is relaxed, and so on. For recent related studies considering extensions to the minimal $\Lambda$CDM model we refer the reader to e.g.~\cite{bellini,joudaki,archidiaconogiusarmamelchiorrimena,feeney,archidiacono1,archidiacono2,mirizzi,verde,gariazzo,archidiacono3,bergstrom,rossi,
Zhang:2015rha,divalentino1,gerbino,divalentino2,zhang,kitching,moresco,canac,archidiacono4,kumar,bouchet,Kumar:2017dnp,Guo:2017hea,Zhang:2017rbg,
Li:2017iur,Yang:2017amu,Feng:2017nss,Dirian:2017pwp,Feng:2017mfs,Lorenz:2017fgo}, as well as Sec.~\ref{extendedparameterspace}. For other recent studies which investigate the effect of systematics or the use of datasets not considered here (e.g. cross-correlations between CMB and large-scale structure) see e.g.~\cite{Couchot:2017pvz,Doux:2017tsv}.

\begin{table}[h!]
\begin{tabular}{|c?c|c|}
\hline
Parameter & Prior & Name\\
\hline
\hline
$\Omega_b h^2$ & [0.005,0.1] & \\
\hline
$\Omega _c h^2$ & [0.01,0.99] & \\
\hline
$\Theta_s$ & [0.5,10] & \\
\hline
$\tau$ & [0.01,0.8] & \\
 & 0.055 $\pm$ 0.009 & $\tau0p055$ \\
\hline
$n_s$ & [0.8,1.2] & \\
\hline
$\log(10^{10}A_s)$ & [2,4] & \\
\hline
$M_\nu$ (eV) & [0,3] & \\
\hline
$H_0$ (km/s/Mpc) & [20,100] & (\textit{Implicit}) \\
 & 72.5 $\pm$ 2.5 & $H072p5$ \\
 & 73.02 $\pm$ 1.79 & $H073p02$ \\
\hline
$1-b$ & [0.1,1.3] & \\
\hline
$b_{\text{HF}}$ & [0,10] & \\
\hline
$P_{\text{HF}}$ & [0,10000] & \\
\hline
\end{tabular}
\caption{Priors on cosmological and nuisance parameters considered in this work. Priors on a parameter $p$ of the form $[A,B]$ are uniform within the range $A<p<B$, whereas priors of the form $A \pm B$ are Gaussian with central value and variance given by $A$ and $B$, respectively. The first seven rows refer to the basic parameter vector in Eq.(\ref{parameters}). $H_0$ refers to the Hubble parameter and is a derived parameter, whereas $1-b$ is the cluster mass bias parameter, see Sec.~\ref{sz}. The parameters $b_{\text{HF}}$ and $P_{\text{HF}}$ are nuisance parameters used to model the galaxy power spectrum, see Eq.~(\ref{biasshot}).}
\label{tab:priors}
\end{table}

\subsection{Model comparison between mass hierarchies}
\label{sec:evidence}

As we discussed previously, several works have argued that reaching an upper bound on $M_\nu$ of order $0.1\,{\rm eV}$ would imply having discarded the IH at some statistical significance. In order to quantify the exclusion limits on the IH, a proper model comparison treatment, thus rigorously taking into account volume effects, is required. Various methods which allow the estimation of the exclusion limits on the IH have been devised in the recent literature, see e.g.~\cite{Hannestad:2016fog,Xu:2016ddc,martinanew}. Here, we will briefly describe the simple but rigorous model comparison method which we will use in our work, proposed by Hannestad and Schwetz in \cite{Hannestad:2016fog}, and based on previous work in \cite{blennow}. The method allows the quantification of the statistical significance at which the IH can be discarded, given the cosmological bounds on $M_\nu$. We refer the reader to the original paper \cite{Hannestad:2016fog} for further details.

Let us again consider the likelihood function ${\cal L}$ of the data $\boldsymbol{x}$ given a set of cosmological parameters $\boldsymbol{\theta}$, the mass of the lightest neutrino $m_0 = m_1 \ [m_3]$ for NH [IH], and the discrete parameter $H$ representing the mass hierarchy, with $H=N \ [I]$ for NH [IH] respectively: ${\cal L}(\boldsymbol{x} \vert \boldsymbol{\theta},m_0,H)$. Then, given the prior(s) on cosmological parameters $p(\boldsymbol{\theta})$, we define the likelihood marginalized over cosmological parameters $\boldsymbol{\theta}$ assuming a mass hierarchy $H$, ${\cal E}_H(m_0)$, as:
\begin{eqnarray}
{\cal E}_H(m_0) \equiv \int d\boldsymbol{\theta} \ {\cal L}(\boldsymbol{x} \vert \boldsymbol{\theta},m_0,H)p(\boldsymbol{\theta}) = {\cal L}(\boldsymbol{x} \vert m_0, H) \, .
\label{marginall}
\end{eqnarray}
Imposing an uniform prior $m_0 \geq 0\,{\rm eV}$ and assuming factorizable priors for the other cosmological parameters it is not hard to show that, as a consequence of Bayes' theorem, the posterior probability of a mass hierarchy $H$ given the data $\boldsymbol{x}$, $p_H \equiv p(H\vert\boldsymbol{x})$, can be obtained as below:
\begin{eqnarray}
\label{evidence}
\hskip -0.6 cm p_H = \frac{p(H)\int_0^{\infty} dm_0 \ {\cal E}_H(m_0)}{p(N)\int_0^{\infty} dm_0 \ {\cal E}_N(m_0)+p(I)\int_0^{\infty} dm_0 \ {\cal E}_I(m_0)} \, ,
\end{eqnarray}
where $p(N)$ and $p(I)$ denote priors on the NH and IH respectively, with $p(N)+p(I)=1$. The posterior odds of NH against IH are then given by $p_N/p_I$, whereas the C.L. at which the IH is disfavored, which we refer to as ${\rm CL}_{\rm IH}$, is given by:
\begin{eqnarray}
{\rm CL}_{\rm IH} = 1-p_I \, .
\label{clih}
\end{eqnarray}

The expression in Eq.~\ref{evidence} is correct as long as the assumed prior on $m_0$ is uniform, and the priors on the other cosmological parameters are factorizable. Different choices of priors on $m_0$ will of course lead to a larger or smaller preference for the NH. As an example, \cite{Simpson:2017qvj} considered the effect of logarithmic priors, showing that this leads to a strong preference for the NH (see, however, \cite{Schwetz:2017fey}).

Another valid possibility, which has not explicitly been considered in the recent literature, is that of performing model comparison between the two neutrino hierarchies by imposing an uniform prior on $M_{\nu}$ instead of $m_0$. In this case, it is easy to show that the posterior odds for NH against IH, $p_N/p_I$, is given by (considering for simplicity the case where NH and IH are assigned equal priors):
\begin{eqnarray}
\frac{p_N}{p_I} \equiv \frac{\int_{0.06\, {\rm eV}}^{\infty}dM_{\nu} \ {\cal E}(M_{\nu})}{\int_{0.10\, {\rm eV}}^{\infty}dM_{\nu} \ {\cal E}(M_{\nu})} \, ,
\label{odds}
\end{eqnarray}
where analogously to Eq.~(\ref{marginall}), we define the marginal likelihood ${\cal E}(M_{\nu})$ as:
\begin{eqnarray}
{\cal E}_H(M_{\nu}) \equiv \int d\boldsymbol{\theta} \ {\cal L}(\boldsymbol{x} \vert \boldsymbol{\theta},M_{\nu},H)p(\boldsymbol{\theta}) = {\cal L}(\boldsymbol{x} \vert M_{\nu}, H) \, .
\end{eqnarray}
It is actually easy to show that in the low-mass region of parameter space currently favoured by cosmological data, i.e. $M_{\nu} \lesssim 0.15\, {\rm eV}$, the posterior odds for NH against IH one obtains by choosing a flat prior on $M_{\nu}$ [Eq.~(\ref{odds})] or a flat prior on $m_0$ [Eq.~(\ref{evidence})] are to very good approximation equal. It is also interesting to note that, as is easily seen from Eq.~(\ref{odds}), cosmological data will always prefer the normal hierarchy over the inverted hierarchy, simply as a consequence of volume effects: that is, the volume of parameter space available to the normal hierarchy ($M_{\nu}>0.06\, {\rm eV}$) is greater than that available to the inverted hierarchy ($M_{\nu}>0.1\, {\rm eV}$). For this reason, the way the prior volume is weighted plays a crucial role in determining the preference for one hierarchy over the other (see discussions in~\cite{Simpson:2017qvj,Schwetz:2017fey}).

In our work, we choose to follow the prescription of~\cite{Hannestad:2016fog} (based on a uniform prior on $m_0$) and hence apply Eq.~(\ref{evidence}) to determine the preference for the normal hierarchy over the inverted one from cosmological data.

\section{Datasets and their sensitivity to $M_\nu$}
\label{sec:datasets}

We present below a detailed description of the datasets used in our analyses and their modeling, discussing their sensitivity to the sum of the active neutrino masses. For clarity, all the denominations of the combinations of datasets we consider are summarized in Tab.~\ref{tab:data}. For plots comparing cosmological observables in the presence or absence of massive neutrinos, we refer the reader to~\cite{Lesgourgues:2006nd,Wong:2011ip,Lesgourgues:2012uu,Abazajian:2013oma,book,Lesgourgues:2014zoa} and especially Fig.~1 of the recent~\cite{Archidiacono:2016lnv}.

\begin{table*}[h!]
\begin{tabular}{|c?c|c|}
\hline
Dataset & Content & References\\
\hline\hline
\textit{base} & \textit{PlanckTT}+\textit{lowP} & \cite{planckcosmological,plancklikelihood}\\
\hline
\textit{basepol} & \textit{PlanckTT}+\textit{lowP}+\textit{highP} & \cite{planckcosmological,plancklikelihood}\\
\hline
$P(k)$ & SDSS-III BOSS DR12 CMASS $P(k)$ & \cite{Gil-Marin:2015sqa}\\
\hline
\textit{BAO} & BAO from 6dFGS BAO, WiggleZ, SDSS-III BOSS DR11 LOWZ & \cite{6dfgs,wigglez,dr11}\\
\hline
\textit{BAOFULL} & BAO from 6dFGS, WiggleZ, SDSS-III BOSS DR11 LOWZ, SDSS-III BOSS DR11 CMASS & \cite{6dfgs,wigglez,dr11}\\
\hline
\textit{basePK} & \textit{base}+$P(k)$+\textit{BAO} & \cite{planckcosmological,plancklikelihood,Gil-Marin:2015sqa,6dfgs,wigglez,dr11}\\
\hline
\textit{basepolPK} & \textit{basepol}+$P(k)$+\textit{BAO} & \cite{planckcosmological,plancklikelihood,Gil-Marin:2015sqa,6dfgs,wigglez,dr11}\\
\hline
\textit{baseBAO} & \textit{base}+\textit{BAOFULL} & \cite{planckcosmological,plancklikelihood,Gil-Marin:2015sqa,6dfgs,wigglez,dr11}\\
\hline
\textit{basepolBAO} & \textit{basepol}+\textit{BAOFULL} & \cite{planckcosmological,plancklikelihood,Gil-Marin:2015sqa,6dfgs,wigglez,dr11}\\
\hline
\textit{SZ} & Planck SZ clusters & \cite{plancksz,plancksz1}\\
\hline
\end{tabular}
\caption{Specific datasets and combinations thereof used in this work, and associated references to work where the data is presented and/or discussed.}
\label{tab:data}
\end{table*}

\subsection{Cosmic Microwave Background}\label{CMB}

Neutrinos leave an imprint on the CMB (both at the background and at the perturbation level) in, at least, five different ways, extensively explored in the literature~\cite{Lesgourgues:2006nd,Wong:2011ip,Lesgourgues:2012uu,
Abazajian:2013oma,book,Lesgourgues:2014zoa,Archidiacono:2016lnv}:
\begin{itemize}
\item By delaying the epoch of matter-radiation equality, massive neutrinos lead to an enhanced early integrated Sachs-Wolfe (EISW) effect~\cite{book}. This effect is due to the time-variation of gravitational potentials which occurs during the radiation-dominated, but not during the matter-dominated era, and leads to an enhancement of the first acoustic peak in particular. Traditionally this has been the most relevant neutrino mass signature as far as CMB data is concerned.
\item Because of the same delay as above, light ($f_{\nu}<0.1$) massive neutrinos actually increase the comoving sound horizon at decoupling $r_s(z_{\text{dec}})$, thus increasing the angular size of the sound horizon at decoupling $\Theta_s$ and shifting all the peaks to lower multipoles $\ell$'s~\cite{Lesgourgues:2006nd}.
\item By suppressing the structure growth on small scales due to their large thermal velocities (see further details later in Sec.~\ref{powerspectrum}), reducing the lensing potential and hence the smearing of the high-$\ell$ multipoles due to gravitational lensing~\cite{Lewis:2006fu}. This is a promising route towards determining both the absolute neutrino mass scale and the neutrino mass hierarchy, see e.g. Ref.~\cite{Hall:2012kg,Ade:2013zuv}, because it probes the matter distribution in the linear regime at higher redshift, and because the unlensed background is precisely understood. CMB lensing suffers from systematics as well, although these tend to be of instrumental origin and hence decrease with higher resolution. In fact, a combination of CMB-S4~\cite{s41,s42,s43} lensing and DESI~\cite{Levi:2013gra,Aghamousa:2016zmz,Aghamousa:2016sne} BAO is expected to achieve an uncertainty on $M_\nu$ of $0.016$~eV~\cite{s41}.
\item Massive neutrinos will also lead to a small change in the diffusion scale, which affects the photon diffusion pattern at high-$\ell$ multipoles~\cite{book}, although again this effect is important only for neutrinos which are non-relativistic at decoupling, i.e. for $M_{\nu}>0.6\,{\rm eV}$.
\item Finally, since the enhancement of the first peak due to the EISW depends, in principle, on the precise epoch of transition to the non-relativistic regime of each neutrino species, that is, on the individual neutrino masses, future CMB-only measurements such as those of~\cite{actpol1,actpol2,spt-3g,sa,so,s41,s42,s43,litebird,core1,core2,core3,pixie} could, although only in a very optimistic scenario, provide some hints to unravel the neutrino mass ordering~\cite{book}. Current data instead has no sensitivity to this effect.~\footnote{The effect is below the $\permil$ level for all multipoles, hence well beyond the reach of Planck. The effect will be below the reach of ground-based Stage-III experiments such as Advanced ACTPol \cite{actpol1,actpol2}, SPT-3G \cite{spt-3g}, the Simons Array \cite{sa} and the Simons Observatory \cite{so}. It will most likely be below the reach of ground-based Stage-IV experiments such as CMB-S4 \cite{s41,s42,s43}, or next-generation satellites such as the proposed LiteBIRD \cite{litebird}, COrE \cite{core1,core2}, and PIXIE \cite{pixie}.}
\end{itemize}
Although all the above effects may suggest that the CMB is exquisitely sensitive to the neutrino mass, in practice, the shape of the CMB anisotropy spectra is governed by several parameters, some of which are degenerate among themselves~\cite{bond,howlett}. We refer the reader to the dedicated study of Ref.~\cite{Archidiacono:2016lnv} (see also \cite{Gerbino:2016sgw}).

To assess the impact of massive neutrinos on the CMB, all characteristic times, scales, and density ratios governing the shape of the CMB anisotropy spectrum should be kept fixed, i.e. keeping $z_{\text{eq}}$ and the angular diameter distance to last-scattering $d_A(z_{\text{dec}})$ fixed. This would result in: a decrease in the late integrated Sachs-Wolfe (LISW) effect, which however is poorly constrained owing to the fact that the relevant multipole range is cosmic variance limited; a modest change in the diffusion damping scale for $M_\nu \gtrsim 0.6$ eV; and finally, a $\Delta C_\ell/C_\ell \sim -(M_\nu/0.1 \ {\rm eV})\%$ depletion of the amplitude of the $C_\ell$'s for $20 \lesssim \ell \lesssim 200$, due to a smaller EISW effect, which also contains a sub-$\permil$ effect due to the individual neutrino masses, essentially impossible to detect.

\subsubsection*{\bf \normalsize Baseline combinations of datasets used, and their definitions, I.}

Measurements of the CMB temperature, polarization, and cross-correlation spectra from the Planck 2015 data release~\cite{planckcosmological,planckproducts} are included. We consider a combination of the high-$\ell$ ($30 \leq \ell \leq 2508$) $TT$ likelihood, as well as the low-$\ell$ ($2 \leq \ell \leq 29$) $TT$ likelihood based on the CMB maps recovered with \texttt{Commander}: we refer to this combination as \textbf{\textit{PlanckTT}}. We furthermore include the Planck polarization data in the low-$\ell$ ($2 \leq \ell \leq 29$) likelihood, referring to it as \textbf{\textit{lowP}}. Our baseline model, consisting of a combination of \textit{PlanckTT} and \textit{lowP}, is referred to as \textbf{\textit{base}}.

In addition to the above, we also consider the high-$\ell$ ($30 \leq \ell \leq 1996$) $EE$ and $TE$ likelihood, which we refer to as \textbf{\textit{highP}}. In order to ease the comparison of our results to those previously presented in the literature, we shall add high-$\ell$ polarization measurements to our baseline model separately, referring to the combination of \textit{base} and \textit{highP} as \textbf{\textit{basepol}}. For the purpose of clarity, we have summarized our nomenclature of datasets and their combinations in Tab.~\ref{tab:data}.

All the measurements described above are analyzed by means of the publicly available Planck likelihoods \cite{plancklikelihood}.~\footnote{\href{http://www.cosmos.esa.int/web/planck/pla}{www.cosmos.esa.int/web/planck/pla}} When considering a prior on the optical depth to reionization $\tau$ we shall only consider the $TT$ likelihood in the multipole range $2 \leq \ell \leq 29$. We do so for avoiding double-counting of information, see Sec.~\ref{tau}. Of course, these likelihoods depend also on a number of nuisance parameters, which should be (and are) marginalized over. These nuisance parameters describe, for instance, residual foreground contamination, calibration, and beam-leakage (see Refs.~\cite{planckcosmological,plancklikelihood}).

CMB measurements have been complemented with additional probes which will help breaking the parameter degeneracies discussed. These additional datasets include large-scale structure probes and direct measurements of the Hubble parameter, and will be described in what follows. We make the conservative choice of not including lensing potential measurements, despite measuring $M_{\nu}$ via lensing potential reconstruction is the expected target of the next-generation CMB experiments. This choice is dictated by the observation that lensing potential measurements via reconstruction through the temperature 4-point function are known to be in tension with the lensing amplitude as constrained by the CMB power spectra through the $A_{\text{lens}}$ parameter \cite{planckcosmological} (see also \cite{Calabrese:2008rt,DiValentino:2015ola,DiValentino:2015bja,Addison:2015wyg} for relevant work).

\subsection{Galaxy power spectrum}\label{powerspectrum}

Once CMB data is used to fix the other cosmological parameters, the galaxy power spectrum could in principle be the most sensitive cosmological probe of massive neutrinos among those exploited here. Sub-eV neutrinos behave as a hot dark matter component with large thermal velocities, clustering only on scales below the neutrino free-streaming wavenumber $k_{\text{fs}}$~\cite{Lesgourgues:2012uu,book}:
\begin{eqnarray}
k_{\text{fs}} \simeq 0.018 \ \Omega_m^{1/2} \left (\frac{M_\nu}{1 \rm eV} \right )^{1/2} \ h \ {\rm Mpc}^{-1} \, .
\end{eqnarray}
On scales below the free-streaming scale (or, correspondingly, for wavenumbers larger than the free-streaming wavenumber), neutrinos cannot cluster as their thermal velocity exceeds the escape velocity of the gravitational potentials on those scales. Conversely, on scales well above the free-streaming scale, neutrinos behave as cold dark matter after the transition to the non-relativistic regime. Massive neutrinos leave their imprint on the galaxy power spectrum in several different ways:
\begin{itemize}
\item For wavenumbers $k>k_{\text{fs}}$, the power spectrum in the linear perturbation regime is subject to a scale-independent reduction by a factor of $(1-f_{\nu})^2$, where $f_{\nu} \equiv \Omega_{\nu}/\Omega_m$ is defined as the ratio of the energy content in neutrinos to that in matter~\cite{book}.
\item In addition, the power-spectrum for wavenumbers $k>k_{\text{fs}}$ is further subject to a scale-dependent step-like suppression, starting at $k_{\text{fs}}$ and saturating at $k \sim 1 \ h \ {\rm Mpc}^{-1}$. This suppression is due to the absence of neutrino perturbations in the total matter power spectrum, ultimately due to the fact that neutrinos do not cluster on scales $k>k_{\text{fs}}$. At $k \sim 1 \ h \ {\rm Mpc}^{-1}$, the suppression reaches a constant amplitude of $\Delta P(k)/P(k) \simeq -10f_{\nu}$~\cite{book} (the amplitude of the suppression is independent of redshift, however see the point below).
\item The growth rate of the dark matter perturbations is reduced from $\delta \propto a$ to $\delta \propto a^{1-\frac{3}{5}f_{\nu}}$, due to the absence of gravitational back-reaction effects from free-streaming neutrinos. The redshift dependence of this suppression implies that this effect could be disentangled from that of a similar suppression in the primordial power spectrum by measuring the galaxy power spectrum at several redshifts, which amounts to measuring the time-dependence of the neutrino mass effect~\cite{book}.
\item On very large scales ($10^{-3} < k < 10^{-2}$), the matter power spectrum is enhanced by the presence of massive neutrinos~\cite{neutrinofootprint}.
\item As in the case of the EISW effect in the CMB, the step-like suppression in the matter power spectrum carries a non-trivial dependence on the individual neutrino masses, as it depends on the time of the transition to the non-relativistic regime for each neutrino mass eigenstate~\cite{Lesgourgues:2004ps,Jimenez:2010ev} ($k_{\text{fs}} \propto m_{\nu_i}^{1/2}$), and thus is in principle extremely sensitive to the neutrino mass hierarchy. However, the effect is very small and very hard to measure, even with the most ambitious next-generation large-scale structure surveys~\cite{Pritchard:2008wy,DeBernardis:2009di,Wagner:2012sw}. Through the same effect, the lensed CMB as well as the lensing potential power spectrum could also be sensitive to the neutrino mass hierarchy.
\end{itemize}

Notice that, in principle, once CMB data is used to fix the other cosmological parameters, the galaxy power spectrum could be the most sensitive probe of neutrino masses. In practice, the potential of this dataset is limited by several effects. Galaxy surveys have access to a region of $k$-space $k_{\min}<k<k_{\max}$ where the step-like suppression effect is neither null nor maximal. The minimum wavenumber accessible is limited both by signal-to-noise ratio and by systematics effects, and is typically of order $k \sim 10^{-2} \ h \ {\rm Mpc}^{-1}$, meaning that the fourth effect outlined above is currently not appreciable. The maximum wavenumber accessible is instead limited by the reliability of the non-linear predictions for the matter power spectrum.

\begin{figure*}[t]
\vspace{-0.1cm}
\centering
\includegraphics[width=1.5\columnwidth]{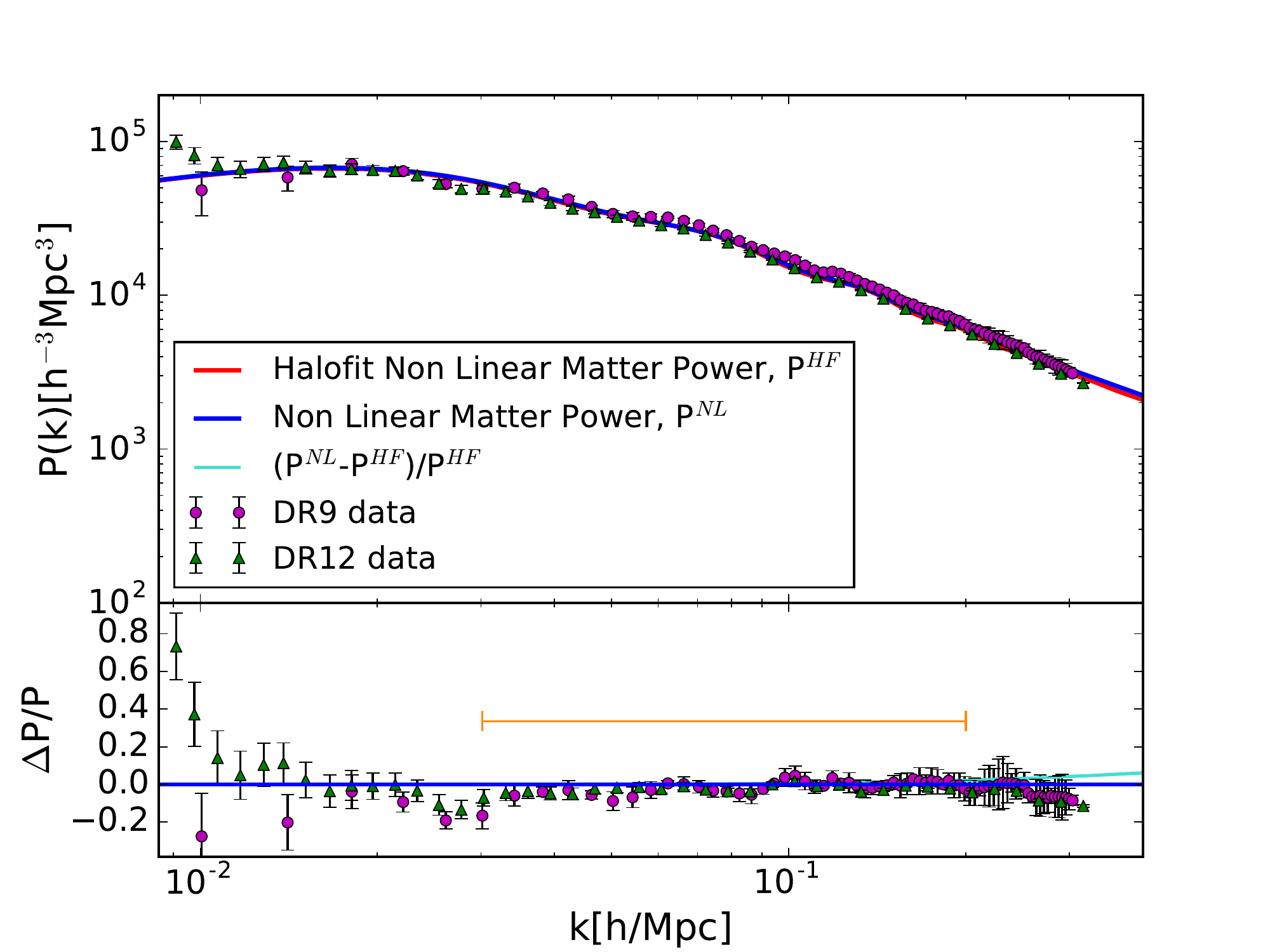}  
\caption{\textit{Top}: Non-linear galaxy power spectrum computed using the \texttt{Halofit} method with the \texttt{camb} code~\cite{Lewis:1999bs} (red line) and the Coyote emulator (blue line)~\cite{Heitmann:2008eq,Heitmann:2013bra,Kwan:2013jva} at z=0.57 for the $\Lambda$CDM best-fit parameters  from \textit{Planck TT} 2015 data and $M_{\nu}=0\,{\rm eV}$ (given that the emulator does not fully implement corrections due to non-zero neutrino masses on small scales). Green triangle data points are the clustering measurements from the BOSS DR12 CMASS sample. The error bars are computed from the diagonal elements $C_{ii}$ of the covariance matrix. For comparison with previous work~\cite{Giusarma:2016phn}, purple circles represent clustering measurements from the BOSS Data Release 9 (DR9) CMASS sample. A very slight suppression in power on small scales (large $k$) of the DR12 sample compared to the DR9 sample is visible. Note that the binning strategy adopted in DR9 and DR12 is different. \textit{Bottom}: Residuals with respect to the non-linear model with \texttt{Halofit}. The orange horizontal line indicates the $k$ range used in our analysis. As it is visually clear, the $k$ range we choose is safe from large non-linear corrections.}
\label{fig:power}
\end{figure*}

At any given redshift, there exists a non-linear wavenumber, above which the galaxy power spectrum is only useful insofar as one is able to model non-linear effects, redshift space distortions, and the possible scale-dependence of the bias (a factor relating the spatial distribution of galaxies and the underlying dark matter density field \cite{kaiserb}) correctly. The non-linear wavenumber depends not only on the redshift of the sample but also on other characteristics of the sample itself (e.g. whether the galaxies are more or less massive). At the present time, the non-linear wavenumber is approximately $k=0.15 \ h \ {\rm Mpc}^{-1}$, whereas for the galaxy sample we will consider (DR12 CMASS, at an effective redshift of $z=0.57$, see footnote~4 for the definition of effective redshift) we will show that wavenumbers smaller than $k=0.2 \ h \ {\rm Mpc}^{-1}$ are safe against large non-linear corrections (see also Fig.~\ref{fig:power}, where the galaxy power spectrum has been evaluated for $M_{\nu}=0\,{\rm eV}$ given that the Coyote emulator adopted~\cite{Heitmann:2008eq,Heitmann:2013bra,Kwan:2013jva} does not fully implement corrections due to non-zero neutrino masses on small scales, and Ref.~\cite{Giusarma:2016phn}).~\footnote{The effective redshift consists of the weighted mean redshift of the galaxies of the sample, with the weights described in \cite{12}.}

The issue of the scale-dependent bias is indeed more subtle than it might seem, given that neutrinos themselves induce a scale-dependent bias~\cite{castorina1,loverde,Raccanelli:2017kht}. A parametrization of the galaxy power spectrum in the presence of massive neutrinos in terms of a scale-independent bias and a shot-noise component [see Eq.(\ref{biasshot})], which in itself adds two extra nuisance parameters, may not capture all the relevant effects at play. Despite these difficulties, the galaxy power spectrum is still a very useful dataset as it helps breaking some of the degeneracies present with CMB-only data, in particular by improving the determination of $\Omega_mh^2$ and $n_s$, the latter being slightly degenerate with $M_\nu$. Moreover, as we shall show in this paper, the galaxy power spectrum represents a conservative dataset (see Sec.~\ref{sec:geometricalandshape}).

Nonetheless, a great deal of effort is being invested into determining the scale-dependent bias from cosmological datasets. There are several promising routes towards achieving this, for instance through CMB lensing, galaxy lensing, cross-correlations of the former with galaxy or quasar clustering measurements, or higher order correlators of the former datasets, see e.g. Refs.~\cite{Gaztanaga:2011yi,Hand:2013xua,Bianchini:2014dla,Pullen:2015vtb,Bianchini:2015yly,Kirk:2015dpw,Pujol:2016lfe,Singh:2016xey,Prat:2016xor}. A sensitivity on $M_\nu$ of $0.023$~eV has been forecasted from a combination of Planck CMB measurements together with weak lensing shear auto-correlation, galaxy auto-correlation, and galaxy-shear cross-correlation from Euclid~\cite{Laureijs:2011gra}, after marginalization over the bias, with the figure improving to $0.01$~eV after including a weak lensing-selected cluster sample from Euclid~\cite{Carbone:2010ik,Joudaki:2011nw,Laureijs:2011gra,Carbone:2011by,Hamann:2012fe,Basse:2013zua}. Similar results are expected to be achieved for certain configurations of the proposed WFIRST survey~\cite{Spergel:2013tha}. It is worth considering that the sensitivity of these datasets would be substantially boosted by determining the scale-dependent bias as discussed above.

A conservative cut-off in wavenumber space, required in order to avoid non-linearities when dealing with galaxy power spectrum data, denies access to the modes where the signature of non-zero $M_{\nu}$ is greatest, i.e. those at high $k$ where the free-streaming suppression effect is most evident. One is then brought to question the usefulness of such data when constraining $M_{\nu}$. Actually, the real power of $P(k)$ rests in its degeneracy breaking ability, when combined with CMB data. For example, $P(k)$ data is extremely useful as far as the determination of certain cosmological parameters  is concerned (e.g. $n_s$, which is degenerate with $M_{\nu}$).

The degeneracy breaking effect of $P(k)$, however, is most evident when in combination with CMB data. As an example, let us consider what is usually referred to as the most significant effect of non-zero $M_{\nu}$ on $P(k)$, that is, a step-like suppression of the small-scale power spectrum. This effect is clearest when one increases $M_{\nu}$ while fixing $(\Omega_m,h)$. However, as we discussed in Sec.~\ref{CMB}, the impact of non-zero $M_{\nu}$ on CMB data is best examined fixing $\Theta_s$. If one adjusts $h$ in order to keep $\Theta_s$ fixed, and in addition keeps $\Omega_bh^2$ and $\Omega_ch^2$ fixed, the power spectrum will be suppressed on both large and small scales, i.e. the result will be a global increase in amplitude~\cite{Poulin:2016nat}. In other words, this reverses the fourth effect listed above. This is just an example of the degeneracy breaking power of $P(k)$ data in combination with CMB data.

Galaxy clustering measurements are addressed by means of the Sloan Digital Sky Survey III (SDSS-III; \cite{sdss}) Baryon Oscillation Spectroscopic Survey (BOSS; \cite{bolton,dawson,smee}) DR12~\cite{alam,12}. The SDSS-III BOSS DR12 CMASS sample covers an effective  volume of $V_{\text{eff}} \approx 7.4 \ {\rm Gpc}^3$~\cite{reid}. It contains 777202 massive galaxies in the range $0.43 < z < 0.7$, at an effective redshift $z = 0.57$ (see footnote~4 for the definition of effective redshift), covering $9376.09$ deg$^2$ over the sky. Here we consider the spherically averaged power spectrum of this sample, as measured by Gil-Mar\'{i}n \textit{et al.} in \cite{Gil-Marin:2015sqa}. We refer to this dataset as $P(k)$. The measured galaxy power spectrum $P_{\text{meas}}^g$ consists of a convolution of the true galaxy power spectrum $P_{\text{true}}^g$ with a window function $W(k_i,k_j)$, which accounts for correlations between the measurements at different scales due to the finite size of the survey geometry:
\begin{eqnarray}
P_{\text{meas}}^g(k_i) = \sum_j W(k_i,k_j)P_{\text{true}}^g(k_j)
\end{eqnarray}
Thus, at each step of the Monte Carlo, we need to convolve the theoretical galaxy power spectrum $P_{\text{th}}$ at the given point in the parameter space with the window function, before comparing it with the measured galaxy power spectrum and constructing the likelihood.

Following previous works \cite{acousticscale,giusarmadeputterhomena}, we model the theoretical galaxy power spectrum as:
\begin{eqnarray}
P_{\text{th}} = b_{\text{HF}}^2P_{\text{HF}\nu}^m(k,z)+P_{\text{HF}}^s \, ,
\label{biasshot}
\end{eqnarray}
where $P_{\text{HF}\nu}^m$ denotes the matter power spectrum calculated at each step by the Boltzmann solver \texttt{camb}, corrected for non-linear effects using the \texttt{Halofit} method~\cite{halofit1,halofit2}. We make use of the modified version of \texttt{Halofit} designed by \cite{birdviel} to improve the treatment of non-linearities in the presence of massive neutrinos. In order to reduce the impact of non-linearities we impose the conservative choice of considering a maximum wavenumber $k_\text{max} = 0.2 \ h \ {\rm Mpc}^{-1}$. As we show in Fig.~\ref{fig:power} (for $M_{\nu}=0\,{\rm eV}$), this region is safe against uncertainties due to non-linear evolution, and is also convenient for comparison with other works which have adopted a similar maximum wavenumber cutoff. The smallest wavenumber we are considering is instead of $k_{\min} = 0.03 \ h \ {\rm Mpc}^{-1}$, and is determined by the control over systematics, which dominate at smaller wavenumbers. The parameters $b_{\text{HF}}$ and $P_{\text{HF}}^s$ denote the scale-independent bias and the shot noise contributions: the former reflects the fact that galaxies are biased tracers of the underlying dark matter distribution, whereas the latter arises from the discrete point-like nature of the galaxies as tracers of the dark matter. We impose flat priors in the range $[0.1,10]$ and $[0,10000]$ respectively for $b_{\text{HF}}$ and $P_{\text{HF}}^s$.

Although in this simple model the bias and shot noise are assumed to be scale-independent, there is no unique prescription for the form of these quantities. In particular, concerning the bias, several theoretically well-motivated scale-dependent functional forms exist in the literature (such as the $Q$ model of \cite{bqq}, that of \cite{amendola}, or that of \cite{Dalal:2007cu} motivated by local primordial non-Gaussianity). It is beyond the scope of our paper to explore the impact of different bias function choices on the neutrino mass bounds. Instead, we simply note that it is not necessarily true that increasing the number of parameters governing the bias shape may result in broader constraints. Indeed, tighter constraints on $M_\nu$ may arise in some of the bias parameterizations with more than one parameter involved, because they might have comparable effects on the power spectrum.

\subsection{Baryon acoustic oscillations}\label{bao}

Prior to the recombination epoch, photons and baryons in the early Universe behave as a tightly coupled fluid, whose evolution is determined by the interplay between the gravitational pull of potential wells, and the restoring force due to the large pressure of the radiation component. The resulting pressure waves which set up, before freezing at recombination, imprint a characteristic scale on the late-time matter clustering, in the form of a localized peak in the two-point correlation function, or a series of smeared peaks in the power spectrum. This scale corresponds to the sound horizon at the drag epoch, denoted by $r_s(z_{\text{drag}})$, where the drag epoch is defined as the time when baryons were released from the Compton drag of photons, see Ref.~\cite{Eisenstein:1997ik}. Then, $r_s(z_{\text{drag}})$ takes the form:
\begin{eqnarray}
r_s(z_{\text{drag}}) = \int_{z_{\text{drag}}}^{\infty} dz \ \frac{c_s(z)}{H(z)} \, ,
\end{eqnarray}
where $c_s(z)$ denotes the sound speed and is given by $c_s(z) = c/\sqrt{3(1+R)}$, with $R=3\rho_b/4\rho_r$ being the ratio of the baryon to photon momentum density. Finally, the baryon drag epoch $z_{\text{drag}}$ is defined as the redshift such that the baryon drag optical depth $\tau_{\text{drag}}$ is equal to one:
\begin{eqnarray}
\tau_{\text{drag}}(\eta_{\text{drag}}) = \frac{4}{3}\frac{\Omega_r}{\Omega_b}\int_0^{z_{\text{drag}}} dz \ \frac{d\eta}{da}\frac{\sigma_T x_e(z)}{1+z} = 1 \, ,
\end{eqnarray}
where $\sigma_T=6.65 \times 10^{-29} \ {\rm m}^2$ denotes the Thomson cross-section and $x_e(z)$ represents the fraction of free electrons.

BAO measurements contain geometrical information in the sense that, as a ``standard ruler'' of known and measured length, they allow for the determination of the angular diameter distance to the redshift of interest, and hence make it possible to map out the expansion history of the Universe after the last scattering. In addition, they are affected by uncertainties due to the non-linear evolution of the matter density field to a lesser extent than the galaxy power spectrum, making them less prone to systematic effects than the latter. An angle-averaged BAO measurement constrains the quantity $D_v(z_{\text{eff}})/r_s(z_{\text{drag}})$, where the dilation scale $D_v$ at the effective redshift of the survey $z_{\text{eff}}$ is a combination of the physical angular diameter distance $D_A(z)$ and the Hubble parameter $H(z)$ (which control the radial and the tangential separations within a given cosmology, respectively):
\begin{eqnarray}
D_v(z) = \left [ (1+z)^2D_A(z)^2\frac{cz}{H(z)} \right ]^{\frac{1}{3}} \, .
\label{dvdilation}
\end{eqnarray}
$D_v$ quantifies the dilation in distances when the fiducial cosmology is modified. The power of the BAO technique resides on its ability of resolving the existing degeneracies present when the CMB data alone is used, in particular in sharpening the determination of $\Omega_m$ and of the Hubble parameter $H_0$, discarding the low values of $H_0$ allowed by the CMB data. 

Massive neutrinos affect both the low-redshift geometry and the growth of structure, and correspondingly BAO measurements. If we increase $M_{\nu}$, while keeping $\Omega_bh^2$ and $\Omega_ch^2$ fixed, the expansion rate at early times is increased, although only for $M_{\nu}>0.6\,{\rm eV}$. Therefore, in order to keep fixed the angular scale of the sound horizon at last scattering $\Theta_s$ (which is very well constrained by the CMB acoustic peak structure), it is necessary to decrease $\Omega_\Lambda$. As $\Omega_\Lambda$ decreases, it is found that $H(z)$ decreases for $z < 1$~\cite{Hou:2011ec,spt}. It can be shown that an increase in $M_{\nu}$ has a negligible effect on $r_s(z_{\text{drag}})$. Hence, we conclude that the main effect of massive neutrinos on BAO measurements is to increase  $D_v(z)/r_s(z_{\text{drag}})$ and decrease $H_0$, as $M_{\nu}$ is increased (see \cite{spt}). It is worth noting that there is no parameter degeneracy which can cancel the effect of a non-zero neutrino mass on BAO data alone, as far as the minimal $\Lambda$CDM+$M_{\nu}$ extended model is concerned~\cite{Archidiacono:2016lnv}.

\subsubsection*{\bf \normalsize Baseline combinations of datasets used, and their definitions, II.}

In this work, we make use of BAO measurements extracted from a number of galaxy surveys. When using BAO measurements in combination with the DR12 CMASS $P(k)$, we consider data from the Six-degree Field Galaxy Survey (6dFGS)~\cite{6dfgs}, the WiggleZ survey~\cite{wigglez}, and the DR11 LOWZ sample~\cite{dr11}, as done in \cite{Giusarma:2016phn}. We refer to the combination of these three BAO measurements as \textbf{\textit{BAO}}. When combining \textit{BAO} with the \textit{base} CMB dataset and the DR12 CMASS $P(k)$ measurements, we refer to the combination as \textbf{\textit{basePK}}. When combining \textit{BAO} with the \textit{basepol} CMB dataset and the DR12 CMASS $P(k)$ measurements, we refer to the combination as \textbf{\textit{basepolPK}}. Recall that we have summarized our nomenclature of datasets (including baseline datasets) and their combinations in Tab.~\ref{tab:data}.

The 6dFGS data consists of a measurement of $r_s(z_{\text{drag}})/D_V(z)$ at $z = 0.106$ (as per the discussion above, $r_s/D_V$ decreases as $M_{\nu}$ is increased). The WiggleZ data instead consist of measurements of the acoustic parameter $A(z)$ at three redshifts: $z = 0.44$, $z = 0.6$, and $z = 0.73$, where the acoustic parameter is defined as:
\begin{eqnarray}
A(z) = \frac{100D_v(z)\sqrt{\Omega_mh^2}}{cz} \, .
\end{eqnarray}
Given the effect of $M_{\nu}$ on $D_v(z)$, $A(z)$ will increase as $M_{\nu}$ increases. Finally, the DR11 LOWZ data consists of a measurement of $D_v(z)/r_s(z_{\text{drag}})$ (which increases as $M_{\nu}$ is increased) at $z = 0.32$. 

Since the BAO feature is measured from the galaxy two-point correlation function, to avoid double counting of information, when considering the \textit{base} and \textit{basepol} datasets we do not include the DR11 CMASS BAO measurements, as the DR11 CMASS and DR12 CMASS volumes overlap. However, if we drop the DR12 CMASS power spectrum from our datasets, we are allowed to add DR11 CMASS BAO measurements without this leading to double-counting of information. Therefore, for completeness, we consider this case as well. Namely, we drop the DR12 CMASS power spectrum from our datasets, replacing it with the DR11 CMASS BAO measurement. This consists of a measurement of $D_v(z_{\text{eff}})/r_s(z_{\text{drag}})$ at $z_{\text{eff}} = 0.57$.

\subsubsection*{\bf \normalsize Baseline combinations of datasets used, and their definitions, III.}

We refer to the combination of the four BAO measurements (6dFGS, WiggleZ, DR11 LOWZ, DR11 CMASS) as \textit{BAOFULL}. We instead refer to the combination of the \textit{base} CMB and the \textbf{\textit{BAOFULL}} datasets with the nomenclature \textbf{\textit{baseBAO}}. When $high-\ell$ polarization CMB data is added to this \textit{baseBAO} dataset, the combination is referred to as \textbf{\textit{basepolBAO}}, see Tab.~\ref{tab:data}. The comparison between \textit{basePK} and \textit{baseBAO}, as well as between \textit{basepolPK} and \textit{basepolBAO}, gives insight into the role played by large-scale structure datasets in constraining neutrino masses. In particular, it allows for an assessment of the relative importance of shape information in the form of the power spectrum against geometrical information in the form of BAO measurements when deriving the neutrino mass bounds. For clarity, all the denominations of the combinations of datasets we consider are summarized in Tab.~\ref{tab:data}.

All the BAO measurements used in this work are tabulated in Tab.~\ref{tab:bao}. Note that we do not include BAO measurements from the DR7 main galaxy sample~\cite{mgs} or from the cross-correlation of DR11 quasars with the Ly$\alpha$ forest absorption \cite{lymana}, and hence our results are not directly comparable to other existing studies which included these measurements.

\begin{table*}[h!]
\begin{tabular}{|c?c|c|c|c|}
\hline
Dataset	& Type of measurement & $z_{\text{eff}}$ & Measurement & Reference \\
\hline\hline
6dFGS & $r_s(z_{\text{drag}})/D_v(z_{\text{eff}})$ & 0.106 & $0.336 \pm 0.015$ & Beutler \textit{et al.}, MNRAS 416 (2011) 3017 \cite{6dfgs} \\
\hline
WiggleZ & $A(z)$ & 0.44 & $0.474 \pm 0.034$ & Blake \textit{et al.}, MNRAS 418 (2011) 1707 \cite{wigglez} \\
 & $A(z)$ & 0.60 & $0.442 \pm 0.020$ & Blake \textit{et al.}, MNRAS 418 (2011) 1707 \cite{wigglez} \\
 & $A(z)$ & 0.73 & $0.424 \pm 0.021$ & Blake \textit{et al.}, MNRAS 418 (2011) 1707 \cite{wigglez} \\
\hline
BOSS DR11 LOWZ & $D_v(z_{\text{eff}})/r_s(z_{\text{drag}})$ & 0.32 & $8.250 \pm 0.170$ & Anderson \textit{et al.}, MNRAS 441 (2014) 1, 24 \cite{dr11} \\
\hline
BOSS DR11 CMASS & $D_v(z_{\text{eff}})/r_s(z_{\text{drag}})$ & 0.57 & $13.773 \pm 0.134$ & Anderson \textit{et al.}, MNRAS 441 (2014) 1, 24 \cite{dr11} \\
\hline
\end{tabular}
\caption{Baryon Acoustic Oscillation measurements considered in this work. From left to right, the columns display the survey, the type of measurement, the effective redshift, the measurement, and the associated reference.}
\label{tab:bao}
\end{table*}

\subsection{Hubble parameter measurements}\label{H0}

Direct measurements of $H_0$ are very important when considering bounds on $M_{\nu}$. With CMB data alone, there exists a strong degeneracy between $M_{\nu}$ and $H_0$ (see e.g. \cite{giusarmadeputtermena}). When $M_{\nu}$ is varied, the distance to last scattering changes as well. Defining $\omega_b \equiv \Omega_bh^2$, $\omega_c \equiv \Omega_ch^2$, $\omega_m \equiv \Omega_mh^2$, $\omega_r \equiv \Omega_rh^2$, $\omega_{\nu} \equiv \Omega_{\nu}h^2$, within a flat Universe, this distance is given by:
\begin{eqnarray}
\chi = c\int_0^{z_{\text{dec}}} \frac{dz}{\sqrt{\omega_r(1+z)^4+\omega_m(1+z)^3+ \left ( 1-\frac{\omega_m}{h^2} \right )}} \, ,
\end{eqnarray}
where $\omega_m = \omega_c + \omega_b + \omega_{\nu}$. The structure of the CMB acoustic peaks leaves little freedom in varying $\omega_c$ and $\omega_b$. Therefore, for what concerns the distance to the last scattering, a change in $M_{\nu}$ can be compensated essentially only by a change in $h$ or, in other words, by a change in $H_0$. This suggests that $M_{\nu}$ and $H_0$ are strongly anti-correlated: the effect on the CMB of increasing $M_{\nu}$ can be easily compensated by a decrease in $H_0$, and vice versa.

In light of the above discussion, we expect a prior on the Hubble parameter to help pinning down the allowed values of $M_{\nu}$ from CMB data. Here, we consider two different priors on the Hubble parameter. The first prior we consider is based on a reanalysis of an older measurement based on the Hubble Space Telescope, the original measurement being $H_0 = (73.8 \pm 2.4) \ {\rm km} \ {\rm s}^{-1}{\rm Mpc}^{-1}$~\cite{riess2011}. The original measurement showed a $\sim 2.4\sigma$ tension with the value of $H_0$ derived from fitting CMB data \cite{Ade:2013zuv,planckcosmological}. The reanalysis, conducted by Efstathiou in Ref.~\cite{revisited}, used the revised geometric maser distance to NGC4258 of Ref.~\cite{ngc4258} as a distance anchor. This reanalysis obtains a more conservative value of $H_0 = (70.6 \pm 3.3) \ {\rm km} \ {\rm s}^{-1}{\rm Mpc}^{-1}$, which agrees with the extracted $H_0$ value from CMB-only within $1\sigma$. We refer to this prior as $H070p6$.

The second prior we consider is based on the most recent HST $2.4\%$ determination of the Hubble parameter in Ref.~\cite{riess2016}. This measurement benefits from more than twice the number of Cepheid variables used to calibrate luminosity distances, with respect to the previous analysis~\cite{riess2011}, as well as from improved determinations of distance anchors. The measured value of the Hubble parameter is $H_0 = (73.02 \pm 1.79) \ {\rm km} \ {\rm s}^{-1}{\rm Mpc}^{-1}$, which is in tension with the CMB-only $H_0$ value by $3\sigma$. We refer to the corresponding prior as $H073p02$.~\footnote{We do not include here the latest $3.8\%$ determination of $H_0$ by the H0LiCOW program. The measurement, based on gravitational time delays of three multiply-imaged quasar systems, yields $H_0 = 71.9^{+2.4}_{-3.0} \ {\rm km} \ {\rm s}^{-1}{\rm Mpc}^{-1}$~\cite{bonvin}.}

A consideration is in order at this point. Given the strong degeneracy between $M_{\nu}$ and $H_0$, we expect the introduction of the two aforementioned priors (especially the $H073p02$ one) to lead to a tighter bound on $M_{\nu}$. At the same time, we expect this bound to be less reliable and/or robust. In other words, such a bound would be quite artificial, as it would be driven by a combination of the tension between direct and primary CMB determinations of $H_0$ and the strong $M_{\nu}-H_0$ degeneracy. We can therefore expect the fit to degrade when any of the two aforementioned priors is introduced. We nonetheless choose to include these prior for a number of reasons. Firstly, the underlying measurement in~\cite{riess2016} has attracted significant attention and hence it is worth assessing its impact on bounds on $M_{\nu}$, subject to the strict caveats we discussed, in light of its potential to break the $M_{\nu}-H_0$ degeneracy. Next, our results including the $H_0$ priors will serve as a warning of the danger of adding datasets which are inconsistent between each other.

\subsection{Optical depth to reionization}\label{tau}

The first generation of galaxies ended the dark ages of the Universe. These galaxies emitted UV photons which gradually ionized the neutral hydrogen which had rendered the Universe transparent following the epoch of recombination, in a process known as reionization (see e.g. Ref.~\cite{inthebeginning} for a review). So far, it is not entirely clear when cosmic reionization took place. Cosmological measurements can constrain the optical depth to reionization $\tau$, which, assuming instantaneous reionization (a very common useful approximation), can be related to the redshift of reionization $z_{\text{re}}$.

Early CMB measurements of $\tau$ from WMAP favored an early-reionization scenario  ($z_\mathrm{re} = 10.6 \pm 1.1$ in the instantaneous reionization approximation \cite{Hinshaw:2012aka}), requiring the presence of sources of reionization at $z\gtrsim 10$. This result was in tension with observations of Ly-$\alpha$ emitters at $z\simeq 7$ (see e.g. \cite{Stark:2010qj,Treu:2013ida,Pentericci:2014nia,Schenker:2014tda,Tilvi:2014oia}), that suggest that reionization ended by $z\simeq 6$. However, the results delivered by the Planck collaboration in the 2015 public data release, using the large-scale (low-$\ell$) polarization observations of the Planck Low Frequency Instrument (LFI) \cite{plancklikelihood} in combination with Planck temperature and lensing data, indicate that $\tau = 0.066 \pm 0.016$ \cite{planckcosmological}, corresponding to a significantly lower value for the redshift of instantaneous reionization: $z_\mathrm{re} = 8.8^{+1.2}_{-1.1}$ (see also \cite{Lattanzi:2016dzq} for an assessment of the role of the cleaning procedure on the lower estimate of $\tau$, and \cite{Meerburg:2017lfh} for an alternative indirect method for measuring large-scale polarization and hence constrain $\tau$ using only small-scale and lensing polarization maps), and thus reducing the need for high-redshift sources of reionization \cite{mesinger,choudhury,Robertson:2015uda,Bouwens:2015vha,mitrachoudhuryferrara}.

The optical depth to reionization is a crucial quantity when considering constraints on the sum of neutrino masses, the reason being that there exist degeneracies between $\tau$ and $M_{\nu}$ (see e.g. \cite{DiValentino:2015sam,Giusarma:2016phn,Archidiacono:2016lnv,howlett,Allison:2015qca,Liu:2015txa,Calabrese:2016eii}). If we consider CMB data only (focusing on the $TT$ spectrum), an increase in $M_{\nu}$, which results in a suppression of structure, reduces the smearing of the damping tail. This effect can be compensated by an increase in $\tau$. Due to the well-known degeneracy between $A_s$ and $\tau$ from CMB temperature data (which is sensitive to the combination $A_se^{-2\tau}$), the value of $A_s$ should also be increased accordingly. However, the value of $A_s$ also determines the overall amplitude of the matter power spectrum, which is furthermore affected by the presence of massive neutrinos, which reduce the small-scale clustering. If, in addition to $TT$ data, low-$\ell$ polarization measurements are considered, the degeneracy between $A_s$ and $\tau$ will be largely alleviated and, consequently, also the multiple ones among the $A_s$, $\tau$, and  $M_{\nu}$ cosmological parameters.

Recently, the Planck collaboration has identified, modeled, and removed previously unaccounted systematic effects in large angular scale polarization data from the Planck High Frequency Instrument (HFI) \cite{plancktau} (see also \cite{planckreionization}). Using the new HFI low-$\ell$ polarization likelihood (that has not been made publicly available by the Planck collaboration), the constraints on $\tau$ have been considerably improved, with a current determination of $\tau = 0.055 \pm 0.009$ \cite{plancktau}, entirely consistent with the value inferred from LFI.

In this work, we explore the impact on the constraints on $M_{\nu}$ of adding a prior on $\tau$. Specifically, we impose a Gaussian prior on the optical depth to reionization of $\tau = 0.055 \pm 0.009$, consistent with the results reported in~\cite{plancktau}. We refer to this prior as $\tau0p055$. We expect this prior to tighten our bounds on $M_{\nu}$. However, a prior on $\tau$ is a proxy for low-$\ell$ polarization spectra (low-$\ell$ $C_\ell^{EE}$, $C_\ell^{BB}$, and $C_\ell^{TE}$). Therefore, as previously stated, when adding a prior on $\tau$, we remove the low-$\ell$ polarization data from our datasets, in order to avoid double-counting information, while keeping low-$\ell$ temperature data.

\subsection{Planck SZ clusters}\label{sz}

The evolution with mass and redshift of galaxy clusters offers a unique probe of both the physical matter density, $\Omega_m$, and the present amplitude of density fluctuations, characterized by the root mean squared of the linear overdensity in spheres of radius $8 \ h^{-1}$Mpc, $\sigma_8$, for a review see e.g. \cite{clusters}. Both quantities are of crucial importance when extracting neutrino mass bounds from large-scale structure, due to the neutrino free-streaming nature.

CMB measurements are able to map galaxy clusters via the Sunyaev-Zeldovich (SZ) effect, which consists of an energy boost to the CMB photons, which are inverse Compton re-scattered by hot electrons (see e.g. \cite{sz1,sz2,sz3}). Therefore, the thermal SZ effect imprints a spectral distortion to CMB photons traveling along the cluster line of sight. The distortion consists of an increase in intensity for frequencies higher than 220 GHz, and a decrease for lower frequencies.

We shall here make us of cluster counts from the latest Planck SZ clusters catalogue, consisting of 439 clusters detected via their SZ signal~\cite{plancksz,plancksz1}. We refer to the dataset as \textit{SZ}. The cluster counts function is given by the number of clusters of a certain mass $M$ within a redshift range $[z,z+dz]$, \textit{i.e.} $dN/dz$:
\begin{eqnarray}
\frac{dN}{dz} \vert_{M > M_{\min}} = f_{\text{sky}}\frac{dV(z)}{dz}\int_{M_{\min}}^{\infty} dM \ \frac{dn}{dM}(M,z) \, .
\end{eqnarray}
The dependence on the underlying cosmological model is encoded in the differential volume $dV/dz$:
\begin{eqnarray}
\frac{dV(z)}{dz} = \frac{4\pi}{H(z)} \int_0^z dz' \ \frac{1}{H^2(z')} \, ,
\end{eqnarray}
through the dependence of the Hubble parameter $H(z)$ on the basic cosmological parameters, and further through the dependence of the cluster mass function $dn/dM$ (calculated through N-body simulations) on the parameters $\Omega_m$ and $\sigma_8$.

The largest source of uncertainty in the interpretation of cluster counts measurements resides in the masses of clusters themselves, which in turn can be inferred by X-ray mass proxies, relying however on the assumption of hydrostatic equilibrium. This assumption can be violated by bulk motion or non-thermal sources of pressure, leading to biases in the derived value of the cluster mass. Further systematics in the X-ray analyses can arise e.g. due to instrument calibration or the temperature structure in the gas. Therefore, it is clear that determinations of cluster masses carry a significant uncertainty, with a typical $\Delta M/M\sim 10-20\%$, quantified via the cluster mass bias parameter, $1-b$:
\begin{eqnarray}
M_X = (1-b)M_{500} \, ,
\end{eqnarray}
where $M_X$ denotes the X-ray extracted cluster mass, and $M_{500}$ the true halo mass, defined as the total mass within a sphere of radius $R_{500}$, $R_{500}$ being the radius within which the mean overdensity of the cluster is 500 times the critical density at that redshift.

As the cluster mass bias $1-b$ is crucial in constraining the values of $\Omega_m$ and $\sigma_8$, and hence the normalization of the matter power spectrum, it plays an important role when constraining $M_{\nu}$. We impose an uniform prior on the cluster mass bias in the range $[0.1,1.3]$, as done in Ref.~\cite{DiValentino:2015sam}, in which it is shown that this choice of $1-b$ leads to the most stringent bounds on the neutrino mass. There exist as well independent lensing measurements of the cluster mass bias, as those provided by the Weighing the Giants project~\cite{wtg}, by the Canadian Cluster Comparison Project~\cite{plancksz1}, and by CMB lensing~\cite{melin} (see also Ref.~\cite{zaldarriaga}). However, we shall not make use of $1-b$ priors based on these independent measurements, as the resulting value of $\sigma_8$ is in slight tension, at the level of 1-2$\sigma$, with primary CMB measurements (however, see \cite{Kitching:2016zkn}).

The value of $\sigma_8$ indicated by weak lensing measurements is smaller than that derived from CMB-only datasets, favoring therefore quite large values of $M_{\nu}$, large enough to suppress the small-scale clustering in a significant way. Therefore, we restrict ourselves to the case in which the cluster mass bias is allowed to freely vary between $0.1$ and $1.3$. It has been shown in \cite{DiValentino:2015sam} that this choice leads to robust and unbiased neutrino mass limits. In this way, the addition of the \textit{SZ} dataset can be considered truly reliable.

\section{Results on $M_\nu$}\label{sec:mnu}

We begin here by analyzing the results obtained for the different datasets and their combinations, assessing their robustness. The constraining power of geometrical versus shape large-scale structure datasets will be discussed in Sec.~\ref{sec:geometricalandshape}. In Sec.~\ref{sec:modelcomparison} we apply the method of \cite{Hannestad:2016fog} and described in Sec.~\ref{sec:evidence} to quantify the exclusion limits on the inverted hierarchy given the bounds on $M_\nu$ presented in the following. The 95\%~C.L. upper bounds on $M_\nu$ we obtain are summarized in Tabs.~\ref{tab:tabmnutt},~\ref{tab:tabmnupol},~\ref{tab:tabmnubao},~\ref{tab:tabmnupolbao}. The C.L.s at which our most constraining datasets disfavor the Inverted Hierarchy, ${\rm CL}_{\rm IH}$, obtained through our analysis in Sec.~\ref{sec:modelcomparison}, are reported in Tab.~\ref{tab:inverted}.

Table~\ref{tab:tabmnutt} shows the results for the more conservative approach when considering CMB data; namely, by neglecting high-$\ell$ polarization data. The limits obtained when the \textit{base} dataset is considered are very close to those quoted in Ref.~\cite{DiValentino:2015sam}, where a three degenerate neutrino spectrum with a lower prior on $M_{\nu}$ of $0.06$~eV was assumed, whereas we have taken a lower prior of $0$~eV. Our choice is driven by the goal of obtaining independent bounds on $M_\nu$ from cosmology alone, making the least amount of assumptions. This different choice of prior is the reason for the (small) discrepancy in our $95\%$~C.L. upper limit on $M_{\nu}$ ($0.716$~eV) and the limit found in Ref.~\cite{DiValentino:2015sam} ($0.754$~eV), and, in general, in all the bounds we shall describe in what follows. That is, these discrepancies are due to differences in the volume of the parameter space explored. When $P(k)$ data are added to the \textit{base}, CMB-only dataset, the neutrino mass limits are considerably improved, reaching $M_{\nu} <0.299$~eV at $95\%$~C.L..

The limits reported in Table~\ref{tab:tabmnutt}, while being consistent with those presented in Ref.~\cite{Giusarma:2016phn} (obtained with an older BOSS full shape power spectrum measurement, the DR9 CMASS $P(k)$), are slightly less constraining. We attribute this mild slight loss of constraining power to the fact that the DR12 $P(k)$ appears slightly suppressed on small scales with respect to the DR9 $P(k)$, see Fig.~\ref{fig:power}. This fact, already noticed for previous data releases, can ultimately be attributed to a very slight change in power following an increase in the mean galaxy density over time due to the tiling (observational) strategy of the survey \cite{cuestapc}. The changes are indeed very small, and the broadband shape of the power spectra for different data releases in fact agree very well within error bars. A small suppression in small-scale power, nonetheless, is expected to favor higher values of $M_\nu$, which help explaining the observed suppression, and this explains the slight difference between our results and those of Ref.~\cite{Giusarma:2016phn}.

While the addition of external datasets, such as a prior on $\tau$ or Planck SZ cluster counts, leads to mild improvements in the constraints on $M_{\nu}$, the tightest bounds are obtained when considering the $H073p02$ prior on the Hubble parameter, due to the large existing degeneracy between $H_0$ and $M_{\nu}$ at the CMB level, and only partly broken via $P(k)$ or BAO measurements. However, as previously discussed, this $H073p02$ measurement shows a significant tension with CMB estimates of the Hubble parameter.~\footnote{See e.g. Refs.~\cite{Poulin:2016nat,h01,h02,h03,h05,h06,h07,h08,h09,h010,h011,h012,h013,h014,h015,h016,h017,h018} for recent works examining this discrepancy and possible solutions.} Therefore, the $95\%$~C.L. limits on $M_{\nu}$ of  $<0.164$, $<0.140$, $<0.136$~eV for the \textit{basePK}+$H073p02$, \textit{basePK}+$H073p02$+$\tau0p055$ and \textit{basePK}+$H073p02$+$\tau0p055$+\textit{SZ} cases should be regarded as the most aggressive limits one can obtain when considering a prior on $H_0$ and neglecting high-$\ell$ polarization data. Indeed, when using the $H070p6$ prior, a less constraining limit of $M_{\nu} <0.219$~eV at $95\%$~C.L. is obtained in the \textit{basePK}+$H070p06$ case, value that is closer to the limits obtained when additional measurements (not related to $H_0$ priors) are added to the \textit{basePK} data combination.

The tension between the $H073p02$ measurement and primary CMB determinations of $H_0$ implies that the very strong bounds obtained using such prior are also the least robust and/or reliable. They are almost entirely driven by the aforementioned tension in combination with the strong $M_{\nu}-H_0$ degeneracy, and hence are somewhat artificial. We expect in fact the quality of the fit to deteriorate in the presence of 2 inconsistent datasets (that is, CMB spectra and $H_0$ prior). To quantify the worsening in fit, we compute the $\Delta \chi^2$ associated to the bestfit, for a given combination of datasets before and after the addition of the $H_0$ prior. For example, for the \textit{basePK} dataset combination, we find $\Delta \chi^2 \equiv \chi^2_{\min}(\textit{basePK}+H073p02)-\chi^2_{\min}(\textit{basePK})=+5.2$, confirming as expected a substantial worsening in fit when the $H073p02$ prior is added to the \textit{basePK} dataset. The above observation reinforces the fact that any bound on $M_{\nu}$ obtained using the $H073p02$ prior should be interpreted with considerable caution, as such bound is most likely artificial.

Table~\ref{tab:tabmnupol} shows the equivalent to Tab.~\ref{tab:tabmnutt} but including high-$\ell$ polarization data. Notice that the limits are considerably tightened. As previously discussed, the tightest bounds are obtained when the $H073p02$ prior is considered. For instance, we obtain $M_{\nu}<0.109$~eV at $95\%$~C.L. from the 
\textit{basepolPK}+$H073p02$+$\tau0p055$ data combination. We caution once more against the very tight bounds obtained with the $H073p02$ being most likely artificial. This is confirmed for example by the $\Delta \chi^2_{\min}=+6.4$ between the \textit{basepolPK}+$H073p02$ and \textit{basepolPK} datasets.

\subsection{Geometric vs shape information}\label{sec:geometricalandshape}

\begin{figure}[b]
\vspace{-0.1cm}
\centering
\includegraphics[width=1.0\columnwidth]{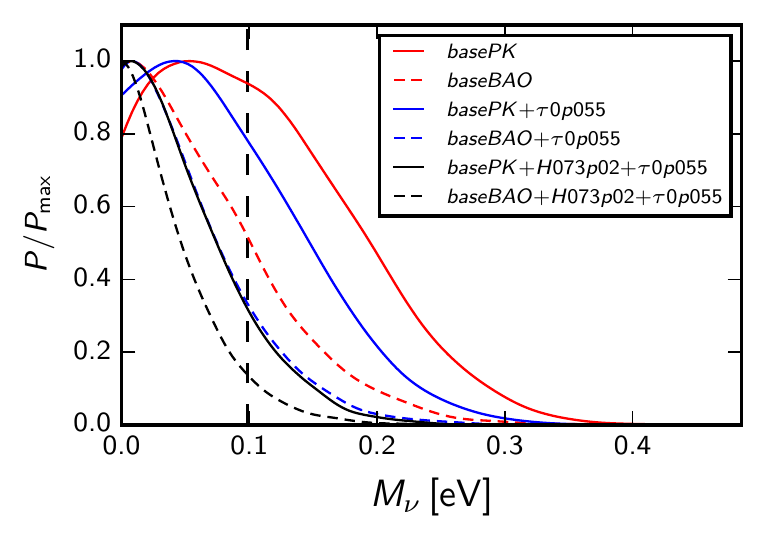}
\caption{Posteriors of $M_\nu$ obtained with baseline datasets \textit{basePK} and \textit{baseBAO}, in combination with additional external datasets. This allows for a comparison of the constraining power of shape information in the form of the full shape galaxy power spectrum, and geometrical information in the form of BAO measurements, when CMB full temperature and low-$\ell$ polarization data are used. To compare the relative constraining power of shape and geometrical information, compare the solid and dashed lines for a given color: red (\textit{basePK} against \textit{baseBAO}), blue (\textit{basePK}+$\tau0p055$ against \textit{baseBAO}+$\tau0p055$), and black (\textit{basePK}+$H073p02+\tau0p055$ against \textit{baseBAO}+$H073p02+\tau0p055$). The dotted line at $M_\nu=0.0986$~eV denotes the minimal allowed mass in the IH scenario. It can be clearly seen that with our current analyses methods geometrical information supersedes shape information in constraining power.}
\label{fig:shape_vs_geometry_planck}
\end{figure}

\begin{figure}[t]
\vspace{-0.1cm}
\centering
\includegraphics[width=1.0\columnwidth]{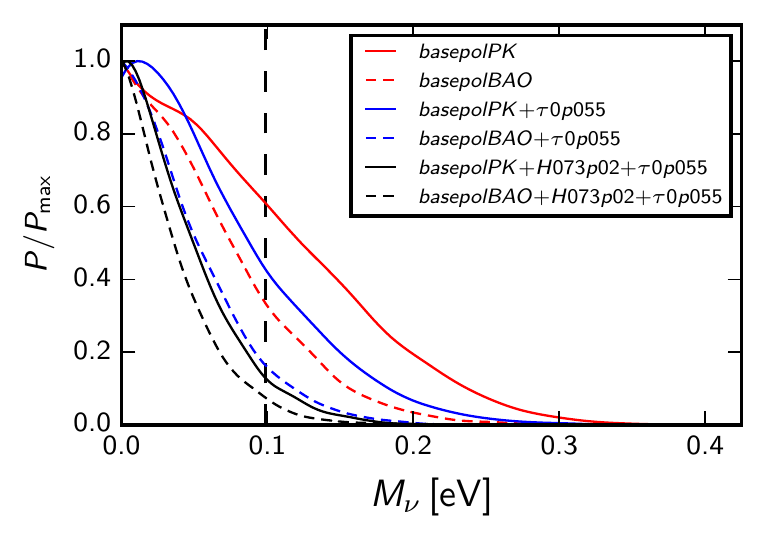}
\caption{As Fig.~\ref{fig:shape_vs_geometry_planck}, but with the addition of high-$\ell$ polarization anisotropy data. Hence, the datasets considered are the baseline datasets \textit{basePK} and \textit{baseBAO}, and combinations with external datasets. Once more, it can be clearly seen that with our current analyses methods geometrical information supersedes shape information in constraining power.}
\label{fig:shape_vs_geometry_planckpol}
\end{figure}

In the following, we shall compare the constraining power of geometrical probes in the form of BAO measurements versus shape probes in the form of power spectrum measurements. For that purpose, we shall replace here the DR12 CMASS $P(k)$ and the \textit{BAO} datasets by the \textit{BAOFULL} dataset, which consists of BAO measurements from the BOSS DR11 (both CMASS and LOWZ samples) survey, the 6dFGS survey, and the WiggleZ survey, see Tab.~\ref{tab:bao} for more details. The main results of this section are summarized in Tabs.~\ref{tab:tabmnubao} and~\ref{tab:tabmnupolbao}, as well as Figs.~\ref{fig:shape_vs_geometry_planck} and~\ref{fig:shape_vs_geometry_planckpol}.
 
Table~\ref{tab:tabmnubao} shows the equivalent to the third, fourth, sixth, eighth and ninth rows of Tab.~\ref{tab:tabmnutt}, but with the shape information from the BOSS DR12 CMASS spectrum replaced by the geometrical BAO information from the BOSS DR11 CMASS measurements. Firstly, we notice that all the geometrical bounds are, in general, much more constraining than the shape bounds, as previously studied and noticed in the literature (see e.g ~\cite{hamannetal,giusarmadeputtermena}, see also \cite{Zhao:2016ecj,Wang:2016tsz} for recent studies on the subject). These studies have shown that, within the minimal $\Lambda$CDM+$M_\nu$ scenario, BAO measurements provide tighter constraints on $M_\nu$ than data from the full power spectrum shape. Nevertheless, it is very important to assess whether these previous findings still hold with the improved statistics and accuracy of today's large-scale structure data (see the recent Ref.~\cite{Archidiacono:2016lnv} for the expectations from future galaxy surveys).

We confirm that this finding still holds with current data. Therefore, current analyses methods of large-scale structure datasets are such that these are still sensitive to massive neutrinos through background rather than perturbation effects, despite the latter are in principle a much more sensitive probe of the effect of massive neutrinos on cosmological observables. However, as we mentioned earlier, this behaviour could be reverted once we are able to determine the amplitude and scale-dependence of the galaxy bias through CMB lensing, cosmic shear, galaxy clustering measurements, and their cross-correlations (see e.g.~\cite{Gaztanaga:2011yi,Hand:2013xua,Bianchini:2014dla,Pullen:2015vtb,Bianchini:2015yly,Kirk:2015dpw,Pujol:2016lfe,Singh:2016xey,Prat:2016xor}). 

Moreover, it is also worth reminding that BAO measurements do include non-linear information through the reconstruction procedure, whereas the same information is prevented from being used in the power spectrum measurements due to the cutoff we imposed at $k=0.2 \ h \ {\rm Mpc}^{-1}$. In order to fully exploit the constraining power of shape measurements, improvements in our analyses methods are necessary: in particular, it is necessary to improve our understanding of the non-linear regime of the galaxy power spectrum in the presence of massive neutrinos, as well as further our understanding of the galaxy bias at a theoretical and observational level.

The addition of shape measurements requires at least two additional nuisance parameters, which in our case are represented by the bias and shot noise parameters. These two parameters relate the measured galaxy power spectrum to the underlying matter power spectrum, the latter being what one can predict once cosmological parameters are known.~\footnote{Moreover, at least another nuisance parameter is required in order to account for systematics in the measured galaxy power spectrum, although the impact of this parameter is almost negligible, as we have checked (see Refs.~\cite{giusarmadeputterhomena,Giusarma:2016phn,cuestapc})}. The prescription we adopted relating the galaxy to the matter power spectrum is among the simplest choices. However, it is not necessarily true that more sophisticated choices with more nuisance parameters would further degrade the constraining power of shape measurements, particularly if we were to obtain a handle on the functional form of the scale-dependent bias~\cite{Gaztanaga:2011yi,Hand:2013xua,Bianchini:2014dla,Pullen:2015vtb,Bianchini:2015yly,Kirk:2015dpw,Pujol:2016lfe,Singh:2016xey,Prat:2016xor}. On the other hand, it remains true that the possibility of benefiting from a large number of modes by increasing the value of $k_{\textrm{max}}$ (which remains one of the factors limiting the constraining power of shape information compared to geometrical one) would require an exquisite knowledge of non-linear corrections, a topic which is the subject of many recent investigations particularly in the scenario where massive neutrinos are present, see e.g. \cite{castorina1,loverde,brandbyge,ichiki,castorina2,costanzi,castorina3,Carbone:2016nzj,zennaro,rizzo}. The conclusion, however, remains that improvements in our current analyses methods, as well as further theoretical and modeling advancements, are necessary to exploit the full constraining power of shape measurements (see also~\cite{Hand:2017ilm,Modi:2017wds,Seljak:2017rmr}).

Finally, we notice that, even without considering the high-$\ell$ polarization data, we obtain the very constraining bound of $M_{\nu}<0.114$~eV at $95\%$~C.L. for the \textit{baseBAO}+$H073p02$+$\tau0p055$+\textit{SZ} datasets. We caution again against the artificialness of bounds obtained using the $H073p02$ prior, as the tension with primary CMB determinations in $H_0$ leads to a degradation in the quality of fit. Nonetheless, even without considering the $H_0$ prior, we still obtain a very constraining bound of $M_\nu<0.151$~eV at $95\%$~C.L. In any case, results adopting these dataset combinations contribute to reinforcing the previous (weak) cosmological hints favouring the NH scenario~\cite{Giusarma:2016phn}.

Table~\ref{tab:tabmnupolbao} shows the equivalent to Tab.~\ref{tab:tabmnubao} but with the high-$\ell$ polarization dataset included, i.e. adding the \textit{highP} Planck dataset in the analyses. We note that the results are quite impressive, and it is interesting to explore how far could one currently get in pushing the neutrino mass limits by means of the most aggressive and least conservative datasets. The tightest limits we find are $M_{\nu}<0.093$~eV at $95\%$~C.L. using the \textit{basepolBAO}+$H073p02$+$\tau0p055$+\textit{SZ} dataset, well below the minimal mass allowed within the IH. Therefore, within the less-conservative approach illustrated here, especially due to the use of the $H073p02$ prior, there exists a weak preference from present cosmological data for a normal hierarchical neutrino mass scheme. Neglecting the information from the $H073p02$ prior, which leads to an artificially tight bound as previously explained, the preference turns out to be weaker ($M_{\nu}<0.118\,{\rm eV}$ from the \textit{basepolBAO}+$\tau0p055$ dataset combination) but still present.

We end with a consideration, stemming from the observation that with our current analyses methods BAO measurements are more constraining than full-shape power spectrum ones. This suggests that, despite uncertainties in the modeling of the galaxy power spectrum due to the unknown absolute scale of the latter (in other words, the size of the bias) and non-linear evolution, the galaxy power spectrum actually represents a conservative dataset given that the bounds on $M_\nu$ obtained using the corresponding BAO dataset are considerably tighter.

In the remainder of the Section we will be concerned with providing a proper quantification of the statistical significance at which we can disfavor the IH, performing a simple but rigorous model comparison analysis.

\subsection{Exclusion limits on the inverted hierarchy}\label{sec:modelcomparison}

Here we apply the method of \cite{Hannestad:2016fog} and described in Sec.~\ref{sec:evidence} to determine the statistical significance at which the inverted hierarchy is disfavored given the bounds on $M_\nu$ just obtained. Our results are summarized in Tab.~\ref{tab:inverted}. In order to quantify the exclusion limits on the inverted hierarchy, we apply Eq.~(\ref{evidence}) to our most constraining dataset combinations, where the criterion for choosing these datasets will be explained below.

Note that in Eq.~(\ref{evidence}) we set $p(N)=p(I)=0.5$. That is, we assign equal priors to NH and IH, which not only is a reasonable choice when considering only cosmological datasets~\cite{Hannestad:2016fog}, but is also the most uninformative and most conservative choice when there is no prior knowledge about the hierarchies. In any case, the formalism we adopt would allow us to introduce informative prior information on the two hierarchies, i.e. $p(N) \neq p(I) \neq 0.5$. It would in this way be possible to include information from oscillation experiments, which suggest a weak preference for the normal hierarchy due to matter effects (see e.g.~\cite{fit1,fit2,fit3,fit4,fit5}). Including this weak preference does not significantly affect our results, precisely because the current sensitivity to the neutrino mass hierarchy from both cosmology and oscillation experiments is extremely weak (see also e.g.~\cite{Hannestad:2016fog}).

We choose to only report the statistical significance at which the IH is discarded for the most constraining dataset combinations, that is, those which disfavor the IH at $>70\%$~C.L.: we have checked that threshold for reaching a $\approx 70\%$~C.L. exclusion limit of the IH is reached by datasets combinations which disfavor at $95\%$~C.L. values of $M_\nu$ greater than $\approx 0.12\,{\rm eV}$. In fact, the most constraining bound within our conservative scheme, obtained through the \textit{baseBAO}+$\tau0p055$ combination (thus disfavoring datasets which exhibit some tension with CMB or galaxy clustering measurements, for a 95\%~C.L. upper limit on $M_\nu$ of $0.151$~eV), falls short of this threshold, and is only able to disfavor the IH at $64\%$~C.L., providing posterior odds for NH versus IH of $1.8:1$.

The hierarchy discrimination is improved when small-scale polarization is added to the aforementioned datasets combination, or when the $H073p02$ prior (and eventually SZ cluster counts) are added to the same datasets combination, leading to a $71\%$~C.L. and $72\%$~C.L. exclusion of the IH respectively. Similar levels of statistical significance for the exclusion of the IH are reached when the datasets combinations \textit{basepolPK}+$H073p02$+$\tau0p055$, \textit{basepolPK}+$H073p02$+$\tau0p055$+\textit{SZ}, and \textit{basepolBAO}+$H073p02$ are considered, leading to $74\%$~C.L., $71\%$~C.L., and $72\%$~C.L. exclusion of the IH respectively. However, it is worth reminding once more that the latter figures relied on the addition of the $H073p02$ prior, which leads to less reliable bounds. It is also worth noting that our most constraining datasets combination(s), that is, \textit{basepolBAO}+$H073p02$+$\tau0p055$(+\textit{SZ}), only provide a $77\%$~C.L. exclusion of the IH.

Our findings are totally consistent with those of \cite{Hannestad:2016fog} and suggest that an improved sensitivity of cosmological datasets is required in order to robustly disfavor the IH, despite current datasets are already able to substantially reduce the volume of parameter space available within this mass ordering. In fact, it has been argued in \cite{Hannestad:2016fog} that a sensitivity of at least $\approx 0.02\,{\rm eV}$ is required in order to provide a $95\%$~C.L. exclusion of the IH. Incidentally, not only does such a sensitivity seem within the reach of post-2020 experiments~\cite{jaffe}, but it would also provide a detection of $M_\nu$ at a significance of at least $3\sigma$, unless non-trivial late-Universe effects are at play (see e.g.~\cite{Beacom:2004yd,bellini}).

\subsection{Bounds on $M_{\nu}$ in extended parameter spaces: a brief discussion}
\label{extendedparameterspace}

Thus far we have explored bounds on $M_{\nu}$ within the assumption of a flat background $\Lambda$CDM cosmology. We have used different dataset combinations, and have identified the \textit{baseBAO} dataset (leading to an upper limit of $M_{\nu}<0.186\, {\rm eV}$) combination as being the one providing one of the strongest bounds while at the same time being one of the most robust to systematics and tensions between datasets.

However, we expect the bounds on $M_{\nu}$ to degrade if we were to open the parameter space: that is, if we were to vary additional parameters other than the 6 base $\Lambda$CDM parameters and $M_{\nu}$. While there is no substantial indication for the need to extend the base set of parameters of the $\Lambda$CDM model (see e.g.~\cite{Raveri:2015maa,Heavens:2017hkr}), one is nonetheless legitimately brought to wonder about the robustness of the obtained bounds against extended parameter spaces.

While a detailed study belongs to a follow-up paper in progress~\cite{inprep}, we nonetheless decide to present two examples of bounds on $M_{\nu}$ within minimally extended parameter spaces. That is, we allow in one case the dark energy equation of state $w$ to vary within the range $[-3,1]$ (parameter space denoted by $\Lambda$CDM+$M_{\nu}$+$w$), and in the other case the curvature energy density $\Omega_k$ to vary freely within the range $[-0.3,0.3]$ (parameter space denoted by $\Lambda$CDM+$M_{\nu}$+$\Omega_k$). Both parameters are known to be relatively strongly degenerate with $M_{\nu}$ and hence we can expect our allowing them to vary to lead to less stringent bounds on $M_{\nu}$. In both cases we consider for simplicity the \textit{baseBAO} dataset, for the reasons described above: therefore, the corresponding bound within the $\Lambda$CDM+$M_{\nu}$ parameter space to which we should compare our results to is $M_{\nu}<0.186 \, {\rm eV}$ at $95\%$~C.L., as reported in the first row of Tab.~\ref{tab:tabmnubao}.

\begin{figure}[!h]
\vspace{-0.1cm}
\centering
\includegraphics[width=1.0\columnwidth]{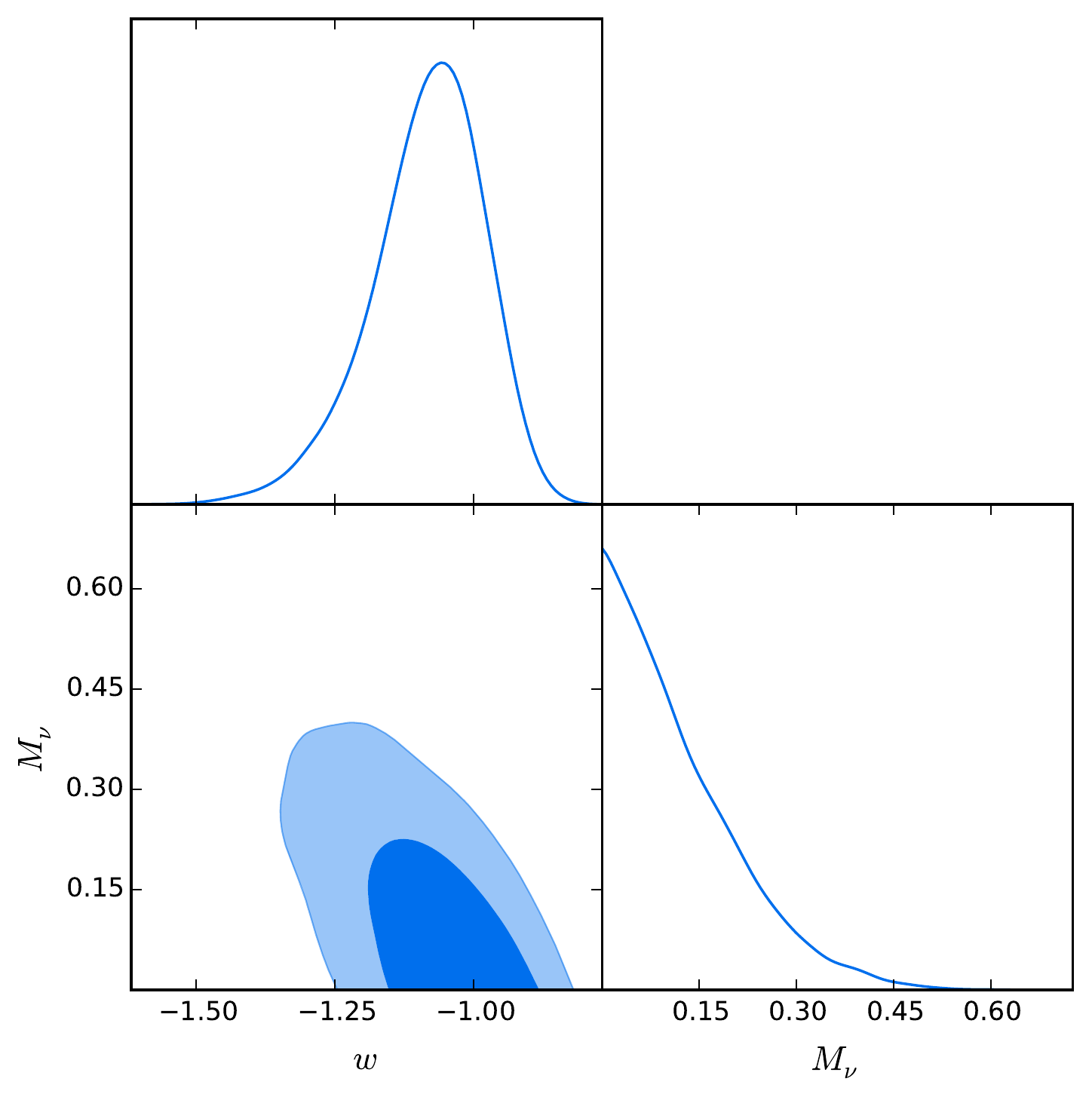}
\caption{$68\%$~C.L. (dark blue) and $95\%$~C.L. (light blue) joint posterior distributions in the $M_{\nu}$-$w$ plane, along with their marginalized posterior distributions, for the \textit{baseBAO} data combination (see the caption of Tab.~\ref{tab:tabmnubao} for further details). Ticks on the $w$-axis of the upper left plot are the same as those for the lower left plot.}
\label{fig:full_w_3mass_tri}
\end{figure}

For the $\Lambda$CDM+$M_{\nu}$+$w$ extension, where we leave the dark energy equation of state $w$ free to vary within the range $[-3,1]$, we can expect the bounds on $M_{\nu}$ to broaden due to a well-known degeneracy between $M_{\nu}$ and $w$~\cite{Hannestad:2005gj}. Specifically, an increase in $M_{\nu}$ can be compensated by a decrease in $w$, due to the mutual degeneracy with $\Omega_m$. Our results confirm this expectation. With the \textit{baseBAO} data combination we find $M_{\nu}<0.313\, {\rm eV}$ at $95\%$~C.L., and $w=-1.08^{+0.09}_{-0.08}$ at $68\%$~C.L., with a correlation coefficient between $M_{\nu}$ and $w$ of $-0.56$.~\footnote{The correlation coefficient between two parameters $i$ and $j$ (in this case $i=M_{\nu}$, $j=w$)is defined as $R=C_{ij}/\sqrt{C_{ii}C_{jj}}$, with $C$ the covariance matrix of cosmological parameters.} The degeneracy between $M_{\nu}$ and $w$ is clearly visible in the triangle plot of Fig.~\ref{fig:full_w_3mass_tri}.

\begin{figure}[!h]
\vspace{-0.1cm}
\centering
\includegraphics[width=1.0\columnwidth]{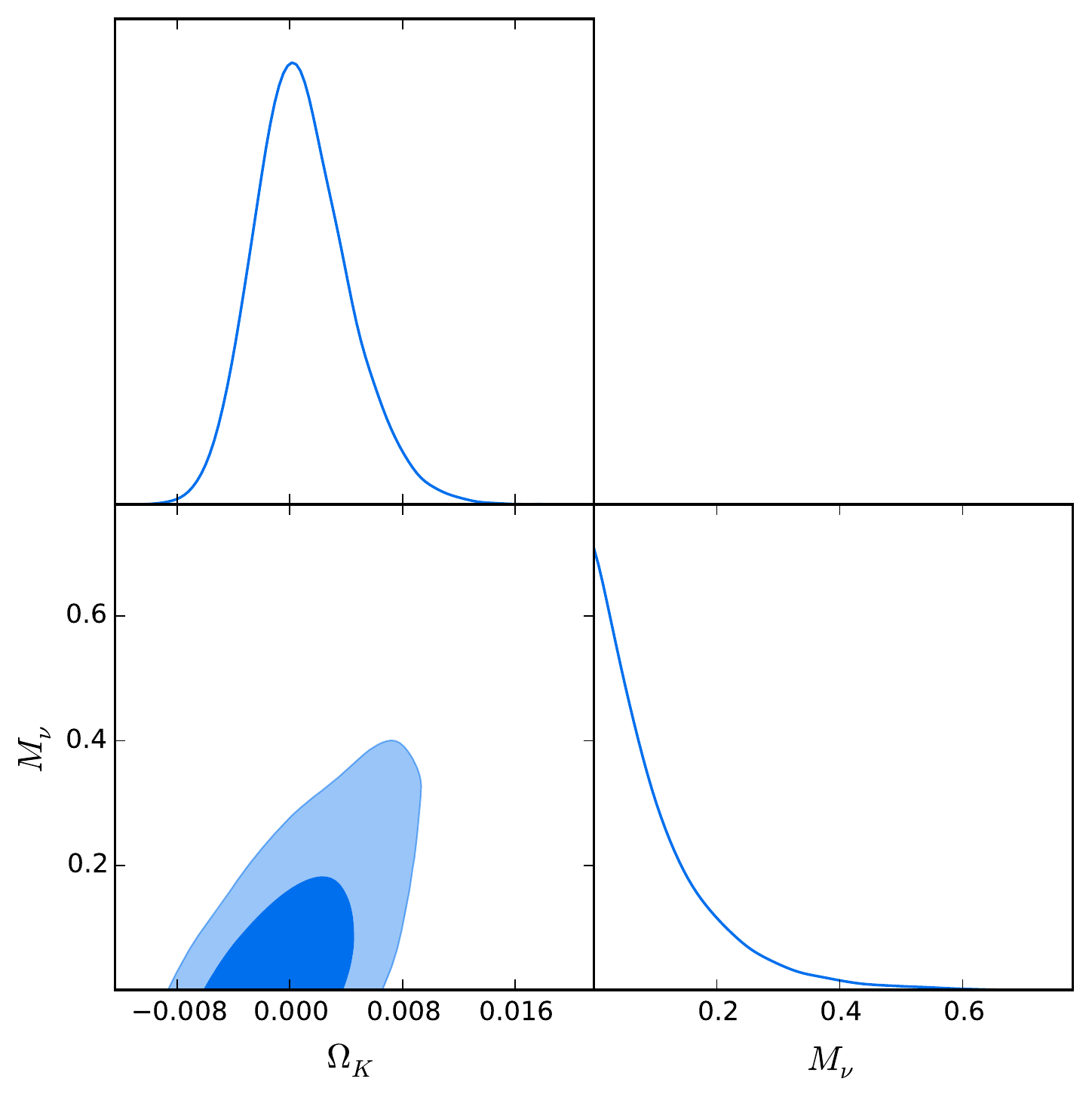}
\caption{$68\%$~C.L. (dark blue) and $95\%$~C.L. (light blue) joint posterior distributions in the $M_{\nu}$-$\Omega_k$ plane, along with their marginalized posterior distributions, for the \textit{baseBAO} data combination (see the caption of Tab.~\ref{tab:tabmnubao} for further details). Ticks on the $\Omega_k$-axis of the upper left plot are the same as those for the lower left plot.}
\label{fig:full_k_3mass_tri}
\end{figure}

For the $\Lambda$CDM+$M_{\nu}$+$\Omega_k$ extension, where we leave the curvature energy density $\Omega_k$ free to vary within the range $[-0.3,0.3]$, we can again expect the bounds on $M_{\nu}$ to broaden due to the three-parameter geometric degeneracy between $h$, $\Omega_{\nu}h^2$ and $\Omega_k$~\cite{howlett}. For the \textit{baseBAO} data combination we find $M_{\nu}<0.299 \, {\rm eV}$ at $95\%$~C.L., and $\Omega_k=0.001^{+0.003}_{-0.004}$ at $68\%$~C.L., with a correlation coefficient between $M_{\nu}$ and $\Omega_k$ of $0.60$. The degeneracy between $M_{\nu}$ and $\Omega_k$ is clearly visible in the triangle plot of Fig.~\ref{fig:full_k_3mass_tri}.

A clarification is in order here: when leaving the dark energy equation of state $w$ and the curvature energy density $\Omega_k$ free to vary, it would be extremely useful to add supernovae data, given that these are extremely sensitive to these two quantities. We have however chosen not to do so in order to ease comparison with the bound $M_{\nu}<0.186\, {\rm eV}$ obtained for the same \textit{baseBAO} combination within the $\Lambda$CDM+$M_{\nu}$ parameter space. Moreover, in this way we are able to reach a conservative conclusion concerning the robustness of $M_{\nu}$ bounds to the $\Lambda$CDM+$M_{\nu}$+$w$ and $\Lambda$CDM+$M_{\nu}$+$\Omega_k$ parameter spaces, as the addition of supernovae data would lead to tighter bounds than the $M_{\nu}<0.313\, {\rm eV}$ and $M_{\nu}<0.299\, {\rm eV}$ quoted.

Of course, as expected, the bounds on $M_{\nu}$ degrade the moment we consider extended parameter spaces. Given our discussion in Sec.~\ref{sec:modelcomparison}, this means within the extended parameter spaces considered the preference for one hierarchy over another essentially vanishes. However, the last statement is not necessarily always true: for instance, in certain models of dynamical dark energy with specific functional forms of $w(z)$, the constraints on $M_{\nu}$ can get tighter: an example is the holographic dark energy model, within which bounds on $M_{\nu}$ have been shown to be substantially tighter than within a $\Lambda$CDM Universe~\cite{Zhang:2015rha,zhang,Wang:2016tsz}. An interesting thing to note, however, is that within better than $1\sigma$ uncertainties (i.e. within $\sim 68\%$~C.L.), both $w$ and $\Omega_k$ are compatible with the values to which they are fixed within the minimal $\Lambda$CDM+$M_{\nu}$ parameter space, that is, $-1$ and $0$ respectively.

\section{Conclusions}\label{sec:conclusion}

Neutrino oscillation experiments provide information on the two mass splittings governing the solar and atmospheric neutrino transitions, but are unable to measure the total neutrino mass scale, $M_{\nu}$. The sign of the largest mass splitting, the atmospheric mass gap, remains unknown. The two resulting possibilities are the so-called normal (positive sign) or inverted (negative sign) mass hierarchies. While in the normal hierarchy scheme neutrino oscillation results set the minimum allowed total neutrino mass $M_{\nu}$ to be approximately equal to $\Mnumin\sim0.06$~eV, in the inverted one this lower limit is $\Mnumin\sim0.1$~eV.

Currently, cosmology provides the tightest bounds on the total neutrino mass $M_{\nu}$, i.e. on the sum of the three active neutrino states. If these cosmological bounds turned out to be robustly and significantly smaller than the minimum allowed in the inverted hierarchy, then one would indeed determine the neutrino mass hierarchy via cosmological measurements. In order to prepare ourselves for the hierarchy extraction, an assessment of the cosmological neutrino mass limits, studying their robustness against different priors and assumptions concerning the neutrino mass distribution among the three neutrino mass eigenstates, is mandatory. Moreover, the development and application of rigorous model comparison methods to assess the preference for one hierarchy over the other is necessary. In this work, we have analyzed some of the most recent publicly available datasets to provide updated constraints on the sum of the three active neutrino masses, $M_\nu$, from cosmology.

One very interesting aspect is whether the information concerning the total neutrino mass from the large-scale structure of the universe in its geometrical form (i.e. via the BAO signature) supersedes that of full-shape measurements of the power spectrum. While previous studies have addressed the question with former galaxy clustering datasets, it is timely to explore the situation with current galaxy catalogs, covering much larger volumes, benefiting from smaller error-bars and also from improved, more accurate descriptions of the mildly non-linear regime in the matter power spectrum.

We find that, despite the latest measurements of the galaxy power spectrum cover a vast volume of our universe, the BAO signature extracted from comparable datasets is still more powerful than the full-shape information, within the minimal $\Lambda$CDM+$M_\nu$ model studied here. This statement is expected to change within the context of extended cosmological models, such as those with non-zero curvature or a time-dependent dark energy equation of state, and we reserve this study to future work~\cite{inprep} (whereas a short discussion on the robustness of the bounds on $M_{\nu}$ within extended parameter spaces is provided in Appendix B).

The reason for the supremacy of BAO measurements over shape information is due to the cutoff in $k$-space imposed when treating the power spectrum. This cutoff is required to avoid the impact of non-linear evolution. It is worth reminding once more that BAO measurements contain non-linear information wrapped in with the reconstruction procedure. This same non-linear information cannot be used in the power spectrum due to the choice of the conservative cutoff in $k$-space. Moreover, the need for at least two additional nuisance parameters relating the galaxy power spectrum to the underlying matter power spectrum further degrades the constraining power of the latter. Therefore, the stronger constraints obtained through geometrical rather than shape measurements should not be seen as a limitation of the constraining power of the latter, rather as a limitation of methods currently used to analyze these datasets. A deeper understanding of the non-linear regime of the galaxy power spectrum in the presence of massive neutrinos, as well as further understanding of the galaxy bias at a theoretical and observational level, are required: it is worth noting that a lot of effort is being invested into tackling these issues.

Finally, in this work we have presented the tightest up-to-date neutrino mass constraints among those which can be found in the literature. Neglecting the debated prior on the Hubble constant of $H_0 = (73.02 \pm 1.79) \ {\rm km} \ {\rm s}^{-1}{\rm Mpc}^{-1}$, the tightest $95\%$~C.L. upper bound we find is $M_\nu<0.151$~eV (assuming a degenerate spectrum), from CMB temperature anisotropies, BAO and $\tau$ measurements. Adding Planck high-$\ell$ polarization data tightens the previous bound to $M_\nu<0.118$~eV. Further improvements are possible if a prior on the Hubble parameter is also added. In this less conservative approach, the $95\%$~C.L. neutrino mass upper limit is brought down to the level of $\sim 0.09$~eV, indicating a weak preference for the normal neutrino hierarchy due to volume effects. Our work also suggests that we can identify a restricted set of conservative but robust datasets: this includes CMB temperature data, as well as BAO measurements and galaxy power spectrum data, after adequate corrections for non-linearities. These datasets allow us to identify a robust upper bound of $\sim 0.15\,{\rm eV}$ on $M_{\nu}$ from cosmological data alone.

In addition to providing updated bounds on the total neutrino mass, we have also performed a simple but robust model comparison analysis, aimed at quantifying the exclusion limits on the inverted hierarchy from current datasets. Our findings indicate that, despite the very stringent upper bounds we have just outlined, current data is not able to conclusively favor the NH over the IH. Within our most conservative scheme, we are able to disfavor the IH with a significance of at most $64\%$~C.L., corresponding to posterior odds of NH over IH of $1.8:1$. Even the most constraining and less conservative datasets combinations are able at most to disfavor the IH at $77\%$~C.L., with posterior odds of NH against IH of $3.3:1$. This suggests that further improvements in sensitivity, down to the level of $0.02\,{\rm eV}$, are required in order for cosmology to conclusively disfavor the IH. Fortunately, it looks like a combination of data from near-future CMB experiments and galaxy surveys should be able to reach this target.

We conclude that our findings, while unable to robustly disfavor the inverted neutrino mass ordering, significantly reduce the volume of parameter space allowed within this mass hierarchy. The more robustly future bounds will be able to disfavor the region of parameter space with $M_\nu > 0.1\,{\rm eV}$, the more the IH will be put under pressure with respect to the NH. In other words future cosmological data, in the absence of a neutrino mass detection, are expected to reinforce the current mild preference for the normal hierarchy mass ordering. On the other hand, if the underlying mass hierarchy is the inverted one, a cosmological detection of the neutrino mass scale could be quick approaching. In any case, we expect neutrino cosmology to remain an active and exciting field of discovery in the upcoming years.

\section*{Appendix A: The \textit{3deg} approximation}

Throughout the paper we have presented bounds within the \textit{3deg} approximation of a neutrino mass spectrum with three massive degenerate mass eigenstates. The choice was motived, as discussed in Sec.~\ref{sec:intro}, by the observations that the NH and IH mass splittings have a tiny effect on cosmological data, when compared to the \textit{3deg} approximation with the same value of the total mass $M_\nu$. Here we discuss the conditions under which this approximation is mathematically speaking valid. We also briefly discuss why the \textit{3deg} approximation is nonetheless physically accurate given the sensitivity of current data.

Mathematically speaking, the \textit{3deg} approximation is valid as long as:
\begin{eqnarray}
m_0 \gg \vert m_i-m_j \vert \quad \, , \quad \forall i,j=1,2,3 \, ,
\end{eqnarray}
where $m_0=m_1 \ [m_3]$ in the NH [IH] scenario (see Sec.~\ref{sec:intro} for the definition of the labeling of the three mass eigenstates). Recall that, according to our convention, $m_1<m_2<m_3 \quad [m_3<m_1<m_2]$ in the NH [IH]. Therefore, the \textit{3deg} approximation is strictly speaking valid when the absolute neutrino mass scale is much larger than the individual mass splittings. A good candidate for a figure of merit to quantify the goodness of the \textit{3deg} approximation can then be obtained by considering the ratio of any given mass difference, over a quantity proportional to the absolute neutrino mass scale. This leads us to consider the following figure(s) of merit:
\begin{eqnarray}
\zeta_{ij} \equiv \frac{3\vert m_i-m_j \vert}{M_\nu} \, ,
\label{zeta}
\end{eqnarray}
where the indices $i,j$ run over $i,j=1,2,3$. The figures of merit $\zeta_{ij}$ quantify the goodness of the \textit{3deg} approximation. In the case where the \textit{3deg} approximation were exact (which, of course, is physically impossible given the non-zero mass-squared splittings), one would have $\zeta_{ij}=0$. The \textit{3deg} approximation, then, can be considered valid from a practical point of view as long as $\zeta_{ij}$ is sufficiently small, where the amount of deviation from $\zeta_{ij}=0$ one can tolerate defines what is sufficiently small and hence the validity criterion for the \textit{3deg} approximation.

\begin{figure}[!h]
\vspace{-0.1cm}
\centering
\includegraphics[width=1.0\columnwidth]{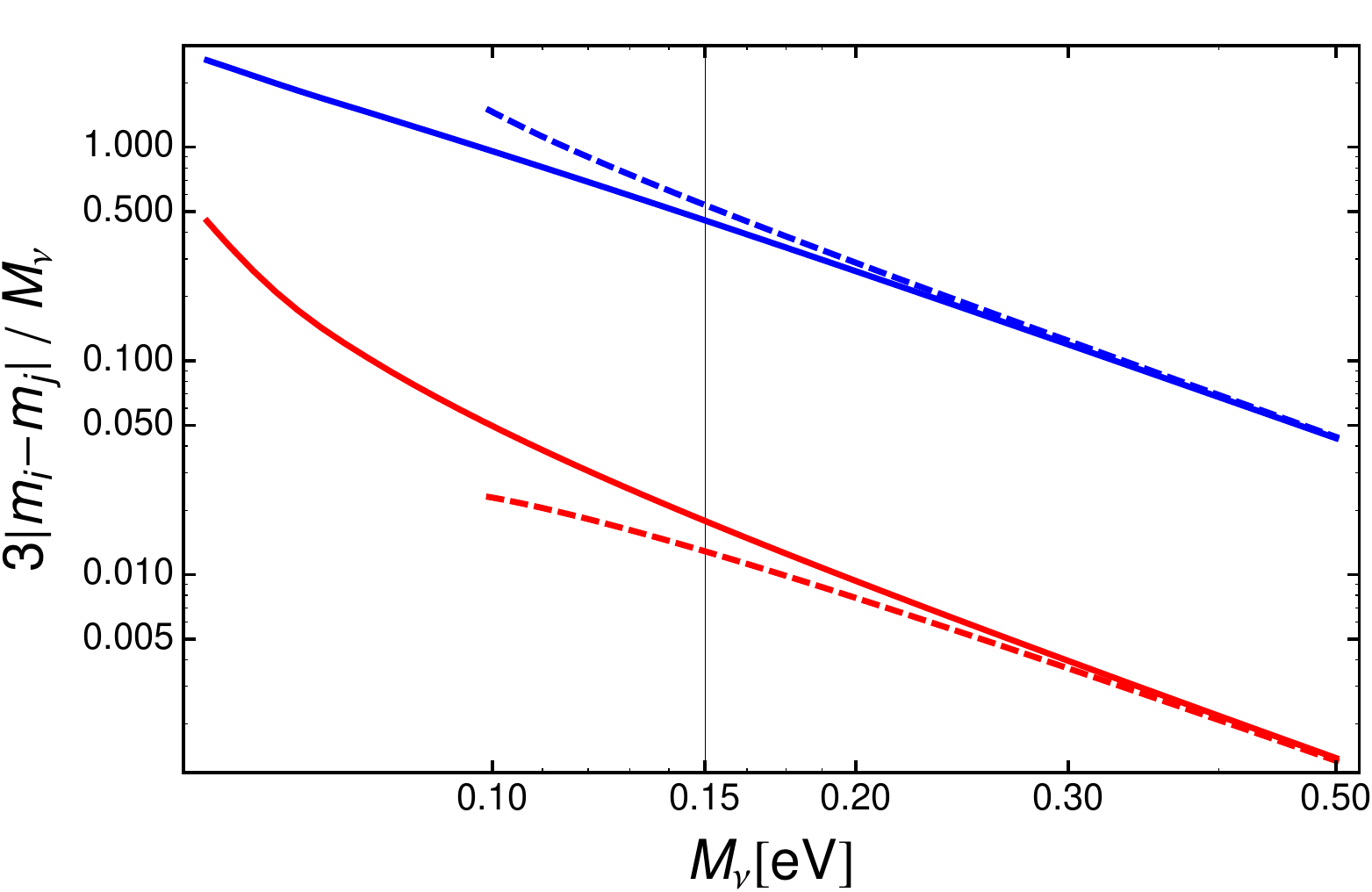}
\caption{Figures of merit $\zeta_{ij}$, defined in Eq.~(\ref{zeta}) and which quantify the goodness of the \textit{3deg} approximation, as a function of the total neutrino mass $M_\nu$. $\zeta_{ij}=0$ (not displayed in this plot due to the logarithmic scale on the $y$ axis) corresponds to the unphysical case where the \textit{3deg} approximation is exact. The red lines correspond to $i,j=1,2$ [that is, $\zeta=3(m_2-m_1)/M_\nu$], whereas the blue lines correspond to $i,j=1,3$ [that is, $\zeta=3\vert m_3-m_1 \vert/M_\nu$], with solid and dashed lines corresponding to the NH and IH scenarios respectively. The solid vertical line at $M_\nu=0.15\,{\rm eV}$ represents the indicative upper limit on $M_\nu$ of $0.15\,{\rm eV}$ obtained in our analysis.}
\label{fig:mass_diff_vs_mnu}
\end{figure}

\begin{figure}[!h]
\vspace{-0.1cm}
\centering
\includegraphics[width=1.0\columnwidth]{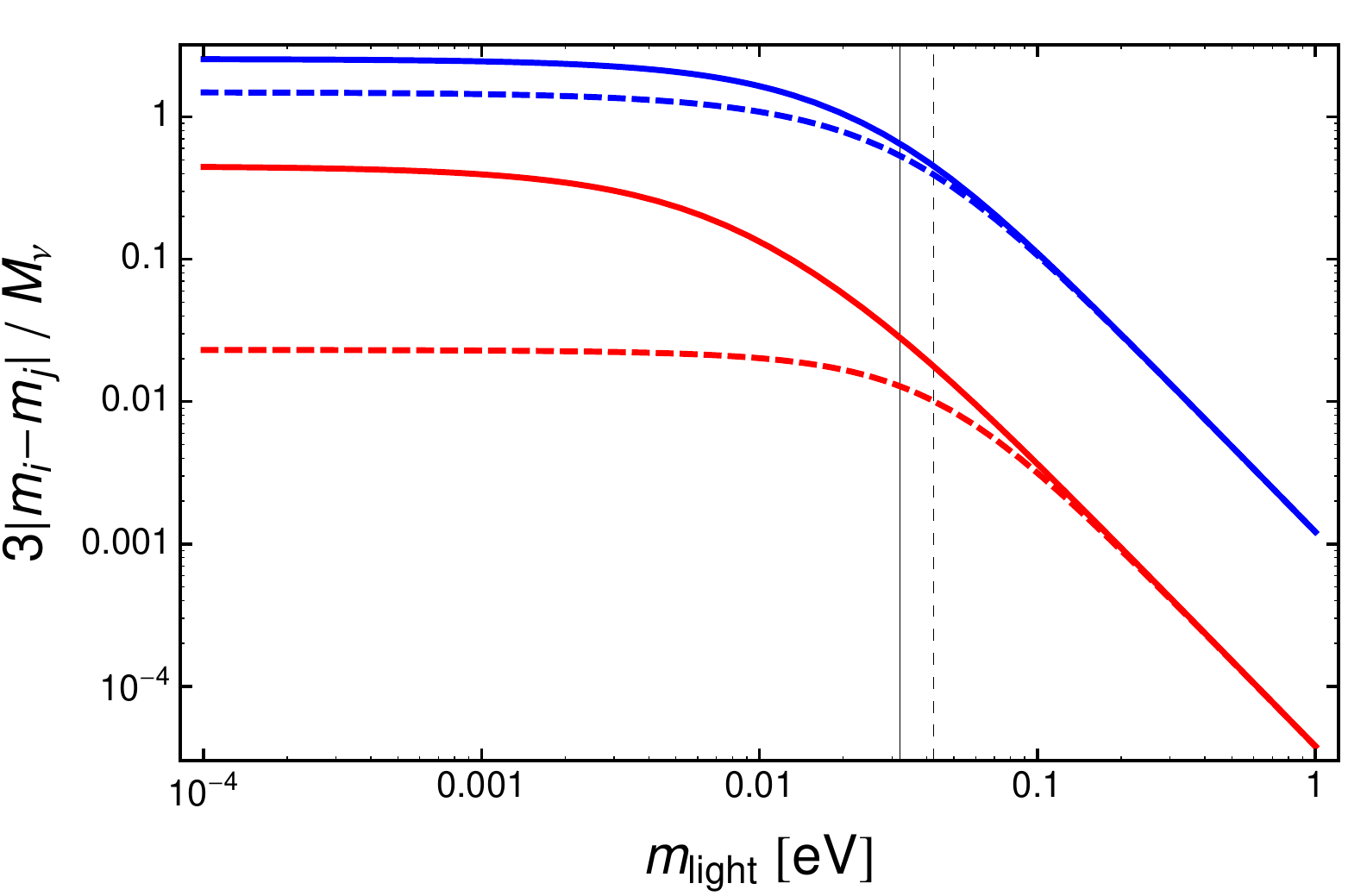}
\caption{As Fig.~\ref{fig:mass_diff_vs_mnu}, but with the figures of merit plotted against the mass of the lightest mass eigenstate $m_0=m_1 \ [m_3]$ for NH [IH]. The solid and dashed vertical lines at $\simeq 0.03\,{\rm eV}$ and $\simeq 0.04\,{\rm eV}$ respectively represent the masses of $m_0$ corresponding to the indicative upper limit on $M_\nu$ of $0.15\,{\rm eV}$ obtained in our analysis.}
\label{fig:mass_diff_vs_m0}
\end{figure}

In Fig.~\ref{fig:mass_diff_vs_mnu} we plot our figure(s) of merit $\zeta_{ij}$, for $i,j=1,2$ (red) and $i,j=1,3$ (blue) in Eq.~(\ref{zeta}) and for the NH (solid) and IH (dashed) scenarios (see the caption for details), against the total neutrino mass $M_\nu$. We plot the same quantities, but this time against the lightest neutrino mass $m_0 = m_1 \ [m_3]$ for the NH [IH], in Fig.~\ref{fig:mass_diff_vs_m0}. As we discussed previously, the \textit{3deg} approximation would be exact if $\zeta_{ij}=0$ (which of course cannot be displayed due to the choice of a logarithmic scale for the $y$ axis).

As we already discussed, the decision of whether or not \textit{3deg} is a sensible approximation mathematically speaking depends on the amount of deviation from $\zeta_{ij}=0$ that can be tolerated. As an example, from Fig.~\ref{fig:mass_diff_vs_mnu} and Fig.~\ref{fig:mass_diff_vs_m0} we see that, considering an indicative value of $M_\nu \approx 0.15\,{\rm eV}$, the value of $\zeta_{13} \approx 0.4$, indicating a $\approx 40\%$ deviation from the exact \textit{3deg} scenario, which can hardly be considered small.

This indicates that, within the remaining allowed region of parameter space, the \textit{3deg} approximation is mathematically speaking not valid. It is worth remarking that there is a degree of residual model dependency as this conclusion was reached taking at face value the indicative upper limit on $M_\nu$ of $\approx 0.15\,{\rm eV}$, which has been derived under the assumption of a flat $\Lambda$CDM background. One can generically expect the bounds we obtained to be loosened to some extent if considering extended cosmological scenarios (although this needs not necessarily always be the case).

A different issue is, instead, whether the \textit{3deg} approximation is physically appropriate, given the sensitivity of current and near-future experiments. The issue has been discussed extensively in the literature, and in particular in some recent works~\cite{Giusarma:2016phn,Gerbino:2016sgw,martinanew}. It has been argued that, if $M_\nu > 0.1\,{\rm eV}$, future cosmological observations, while measuring $M_\nu$ with high accuracy, will not be able to discriminate between the NH and the IH. In any case, cosmological measurements in combination with laboratory experiments will in this case ($M_\nu > 0.1\,{\rm eV}$) play a key role in unravelling the hierarchy~\cite{gerbino}. If $M_\nu < 0.1\,{\rm eV}$, most of the discriminatory power in cosmological data between the NH and the IH is essentially due to volume effects: i.e., the fact that oscillation data force $\Mnumin\simeq0.1\,\eV$ in the IH, implying that the IH has access to a reduced volume of parameter space with respect to the NH.

Another example of the goodness of the \textit{3deg} approximation is provided in~\cite{core2} considering a combination of forecasts for COrE, Euclid, and DESI data. Specifically, \cite{core2} considered a fiducial mock dataset generated implementing the full NH or IH, and then studied whether fitting the fiducial dataset using the \textit{3deg} approximation rather than the ``true" NH or IH would lead to substantial biases. The findings suggest that, apart from small ${\cal O}(0.1\sigma)$ reconstruction biases (which can be removed for $M_{\nu}<0.1\,{\rm eV}$), the \textit{3deg} approximation is able to recover the fiducial value of $M_{\nu}$ (as long as the free parameter is taken to be consistently either $M_{\nu}$ or $m_0$). This suggests that even with near-future cosmological data the \textit{3deg} approximation will still be sufficiently accurate for the purpose of estimating cosmological parameters, and further validates the goodness of the \textit{3deg} approximation in our work.

The conclusion is that current cosmological datasets are sensitive to the total neutrino mass $M_\nu$ rather than to the individual masses $m_i$, implying that the \textit{3deg} approximation is sufficiently precise for the purpose of obtaining reliable cosmological neutrino mass bounds for the time being. On the other hand, for future high precision cosmological data, which could benefit from increased sensitivity and could reliably have access to non-linear scales of the matter power spectrum, modelling the mass splittings correctly will matter.

In conclusion, although the \textit{3deg} approximation is not, mathematically speaking, valid in the remaining volume of parameter space, it is physically speaking a good approximation given the sensitivity of current datasets. However, quantitative claims about disfavoring the inverted hierarchy have to be drawn with care, making use of rigorous model comparison methods.

\section*{Appendix B: The \textit{1mass} approximation}

As argued in a number of works, the ability to robustly reach an upper bound on $M_{\nu}$ of $\approx 0.1\,{\rm eV}$ translates more or less directly into the ability of excluding the inverted hierarchy at a certain statistical significance, as we quantified in Sec.~\ref{sec:modelcomparison}. In this case it is desirable to check whether one's conclusions are affected by assumptions on the underlying neutrino mass spectrum. Throughout our paper we have presented bounds on $M_{\nu}$ making the assumption of a spectrum of three massive degenerate neutrinos, denoted \textit{3deg}. As we have argued extensively (see e.g. Appendix A), given the sensitivity of current data, this assumption does not to any significant extent influence the resulting bounds. Nonetheless, it is interesting and timely to investigate the dependence of neutrino mass bounds under assumptions of different mass spectra, which was recently partly done in~\cite{Giusarma:2016phn}.

Here, as in~\cite{Giusarma:2016phn}, we consider (in addition to the \textit{3deg} spectrum) the approximation spectrum featuring a single massive eigenstate carrying the total mass $M_{\nu}$ together with two massless species. We refer to this scheme by the name \textit{1mass}:
\begin{eqnarray}
m_1 = m_2 = 0 \, , m_3 = M_\nu \quad (\textbf{\textit{1mass}}) \, .
\end{eqnarray}
The motivation for the \textit{1mass} choice is twofold: i) it is the usual approximation adopted when performing cosmological analyses with the total neutrino mass fixed to $\Mnumin=0.06\,\eV$, in order to mimic the minimal mass scenario in the case of the NH ($m_1=0\,\eV$, $m_2\ll m_3$), and ii) it might provide a better description of the underlying neutrino mass ordering in the  $M_\nu<0.1\,\eV$ mass region, in which $m_1\sim m_2\ll m_3$, although a complete assessment goes beyond the scope of our work. The latter is the main motivation for exploring the \textit{1mass} approximation further, given the recent weak cosmological hints favoring the NH.

Before proceeding, it is useful to clarify why we have chosen to focus on results within the \textit{3deg} scheme. As we discussed, it has been observed that the impact of the NH and IH mass splittings on cosmological data is tiny if one compares the \textit{3deg} approximation to the corresponding NH and IH models with the same value of the total mass $M_\nu$. However, this does not necessarily hold when the comparison is made between \textit{3deg} and \textit{1mass}, because the latter always has two pure dark radiation components (see footnote~8 for a definition of dark radiation) throughout the whole expansion history and, in particular, at the present time (on the other hand, NH and IH can have at most one pure radiation component at present time, a situation which occurs in the minimal mass scenario when $m_0 = 0\, {\rm eV}$ and thus only for one specific point in neutrino mass parameter space)~\footnote{Dark radiation consists of any weakly or non-interacting extra radiation component of the Universe, see e.g.~\cite{Archidiacono:2013fha} for a review and \cite{Banerjee:2016suz,Brust:2017nmv} for recent relevant work in connection to neutrino physics. For example, sterile neutrinos may in some models have contributed as dark radiation, see e.g.~\cite{DiValentino:2013qma,Roland:2016gli}, or possibly thermally produced cosmological axions~\cite{Melchiorri:2007cd,Conlon:2013isa}. Dark radiation might also arise in dark sectors with additional relativistic degrees of freedom which decouple from the Standard Model as, for instance, hidden photons (see e.g. \cite{Ackerman:mha,Kaplan:2009de,Cline:2012is,CyrRacine:2012fz,Fan:2013yva,Vogel:2013raa,Petraki:2014uza,Foot:2014uba,
Foot:2014osa,Chacko:2015noa,Foot:2016wvj,Boddy:2016bbu}).}. The extra massless component(s) present in the \textit{1mass} case, but not in the NH and IH (\textit{1mass} features only one extra component compared to the NH and IH if these happen to correspond to the minimal mass scenario where $m_0 = 0\, {\rm eV}$; if $m_0 \neq 0\, {\rm eV}$, \textit{1mass} possesses two extra massless components), are known to have a non-negligible impact on cosmological observables, in particular the CMB anisotropy spectra~\cite{Giusarma:2016phn,Archidiacono:2016lnv}.

Let us now discuss how the bounds on $M_{\nu}$ change when passing from the \textit{3deg} to the \textit{1mass} approximation. We observe that when considering the \textit{base} dataset combinations, and extensions thereof (i.e. the combinations considered in Tab.~\ref{tab:tabmnutt}, where we report the \textit{3deg} results), the bounds obtained within the \textit{1mass} approximation are typically more constraining than the \textit{3deg} ones, by about $\sim 2-8\%$. For example, the $95\%$~C.L. upper bound on $M_{\nu}$ is tightened from $0.716\,{\rm eV}$ to $0.658\,{\rm eV}$ for the \textit{base} combination, from $0.299\,{\rm eV}$ to $0.293\,{\rm eV}$ for the \textit{base}+\textit{P(k)} combination, and from $0.246\,{\rm eV}$ to $0.234\,{\rm eV}$ for the \textit{basePK} combination. When small-scale polarization data is added (see Tab.~\ref{tab:tabmnupol} for the \textit{3deg results}), we observe a reversal in this behaviour: that is, the bounds obtained within the \textit{1mass} approximation are looser than the \textit{3deg} ones. For example, the $95\%$~C.L. upper bound on $M_{\nu}$ is loosened from $0.485\,{\rm eV}$ to $0.619\,{\rm eV}$ for the \textit{basepol} combination, from $0.275\,{\rm eV}$ to $0.300\,{\rm eV}$ for the \textit{basepol}+\textit{P(k)} combination, and from $0.215\,{\rm eV}$ to $0.228\,{\rm eV}$ for the \textit{basepolPK} combination.

Regarding the \textit{baseBAO} and \textit{basepolBAO} dataset combinations and extensions thereof (see Tabs.~\ref{tab:tabmnubao},~\ref{tab:tabmnupolbao} for the \textit{3deg} results), no clear trend emerges when passing from the \textit{3deg} to the \textit{1mass} approximation, although we note that the bounds typically degrade slightly: for example, the $95\%$~C.L. upper bound on $M_{\nu}$ is loosened from $0.186\,{\rm eV}$ to $0.203\,{\rm eV}$ for the \textit{baseBAO} combination, and from $0.153\,{\rm eV}$ to $0.155\,{\rm eV}$ for the \textit{basepolBAO} combination.

We choose not to further investigate the reason behind these tiny but noticeable shifts because, as previously stated, the \textit{1mass} distribution is less ``physical", owing to the presence of two unphysical dark radiation states. Instead, we report these numbers in the interest of noticing how these shifts suggest that, at present, cosmological measurements are starting to become sensitive (albeit in a very weak manner) to the late-time hot dark matter versus dark radiation distribution among the neutrino mass eigenstates, a conclusion which had already been reached in~\cite{Giusarma:2016phn}.

One of the reasons underlying the choice of studying the \textit{1mass} approximation is that this scheme might represent an useful approximation to the minimal mass scenario in the NH. Of course, the possibility that the underlying neutrino hierarchy is inverted is far from being excluded. This raises the question of whether an analogous scheme, which we refer to as \textit{2mass} (already studied in~\cite{Giusarma:2016phn}), might instead approximate the minimal mass scenario in the IH:
\begin{eqnarray}
m_3 = 0 \, , m_1 = m_2 = M_\nu/2 \quad (\textbf{\textit{2mass}}) \, .
\end{eqnarray}
Of course, the previously discussed considerations concerning the non-physicality of the \textit{1mass} approximation (due to the presence of extra pure radiation components) automatically apply to the \textit{2mass} approximation as well. Moreover, we note that bounds on $M_{\nu}$ obtained within the \textit{2mass} approximation (which features one pure radiation state) are always intermediate between those of the \textit{3deg} (which features no pure radiation state) and the \textit{1mass} (which features two pure radiation states) ones (see also e.g.~\cite{Giusarma:2016phn}). This confirms once more that the discrepancy between bounds within these three different approximations are to be attributed to the impact of the unphysical pure radiation states on cosmological observables, in particular the CMB anisotropy spectra. In conclusion, we remark once more that, while the \textit{3deg} approximation is sufficiently accurate given the precision of current data, other approximations which introduce non-physical pure radiation states, such as the \textit{1mass} and \textit{2mass} ones, are not. Adopting these to obtain bounds on $M_{\nu}$ might instead lead to unphysical shifts in the determination of cosmological parameters, and hence should be avoided.

\begin{table*}[h!]
\begin{tabular}{|c?c|}
\hline
Dataset	& $M_\nu$ (95\%~C.L.)\\

\hline\hline
\textit{base} $\equiv$ \textit{Planck TT}+\textit{lowP} & $< 0.716$ eV \\	
\textit{base}+$P(k)$ & $< 0.299$ eV \\
\textit{basePK} $\equiv$ \textit{base}+$P(k)$+\textit{BAO} & $< 0.246$ eV \\
\textit{basePK}+$\tau0p055$ & $< 0.205$ eV \\
\textit{basePK}+\textit{SZ} & $< 0.239$ eV \\
\textit{basePK}+$H073p02$ & $< 0.164$ eV \\
\textit{basePK}+$H070p6$ & $< 0.219$ eV \\
\textit{basePK}+$H073p02$+$\tau0p055$ & $< 0.140$ eV\\
\textit{basePK}+$H073p02$+$\tau0p055$+\textit{SZ} & $< 0.136$ eV\\
\hline
\end{tabular}
\caption{95\%~C.L. upper bounds on the sum of the three active neutrino masses $M_\nu$. The left column lists the combination of cosmological datasets adopted. \textit{PlanckTT} and \textit{lowP} denote measurements of the CMB full temperature and of the low-$\ell$ polarization anisotropies from the Planck satellite 2015 data release, respectively. $P(k)$ denotes the galaxy power spectrum of the CMASS sample from the SDSS-BOSS data release 12 (DR12), with marginalization over the bias and the shot noise, see Eq.~(\ref{biasshot}). \textit{BAO} refers to the combination of BAO measurements from the BOSS data release 11 LOWZ sample, the 6dFGS survey, and the WiggleZ survey (see Table~\ref{tab:bao}). $\tau0p055$ denotes a prior on the optical depth to reionization of $\tau = 0.055 \pm 0.009$ as measured by the Planck HFI. $H073p02$ and $H070p6$ denote priors on the Hubble parameter of $H_0 = 73.02 \pm 1.79 \ {\rm km \ s^{-1} \ Mpc^{-1}}$ and $H_0 = 70.6 \pm 3.3 \ {\rm km \ s^{-1} \ Mpc^{-1}}$, respectively, based on two different HST data analyses. SZ consists of Planck cluster counts measurements via thermal Sunyaev-Zeldovich effects. The right column shows the results (95\%~C.L. upper bounds on $M_\nu$) obtained assuming a degenerate (\textit{3deg}) mass spectrum.}
\label{tab:tabmnutt}
\end{table*}

\begin{table*}[h!]
\begin{tabular}{|c?c|}
\hline
Dataset	& $M_\nu$ (95\%~C.L.)\\

\hline\hline
\textit{basepol} $\equiv$ \textit{PlanckTT}+\textit{lowP}+\textit{highP} & $< 0.485$ eV \\	
\textit{basepol}+$P(k)$ & $< 0.275$ eV \\
\textit{basepolPK}$\equiv$\textit{basepol}+$P(k)$+\textit{BAO} & $< 0.215$ eV \\
\textit{basepolPK}+$\tau0p055$ & $< 0.177$ eV \\
\textit{basepolPK}+\textit{SZ} & $< 0.208$ eV \\
\textit{basepolPK}+$H073p02$ & $< 0.132$ eV \\
\textit{basepolPK}+$H070p6$ & $< 0.196$ eV \\
\textit{basepolPK}+$H073p02$+$\tau0p055$ & $< 0.109$ eV \\
\textit{basepolPK}+$H073p02$+$\tau0p055$+\textit{SZ} & $< 0.117$ eV \\
\hline
\end{tabular}
\caption{As Tab.~\ref{tab:tabmnutt}, but with the addition of \textit{highP}, referring to the small-scale CMB polarization anisotropies data.}
\label{tab:tabmnupol}
\end{table*}

\begin{table*}[h!]
\begin{tabular}{|c?c|}
\hline
Dataset	& $M_\nu$ (95\%~C.L.)\\

\hline\hline
\textit{baseBAO} $\equiv$ \textit{PlanckTT}+\textit{lowP}+\textit{BAOFULL} & $< 0.186$ eV \\	
\textit{baseBAO}+$\tau0p055$ & $< 0.151$ eV \\	
\textit{baseBAO}+$H073p02$ & $< 0.148$ eV \\
\textit{baseBAO}+$H073p02$+$\tau0p055$ & $< 0.115$ eV \\
\textit{baseBAO}+$H073p02$+$\tau0p055$+\textit{SZ} & $< 0.114$ eV \\
\hline
\end{tabular}
\caption{As Tab.~\ref{tab:tabmnutt}, but with the $P(k)$ and the \textit{BAO} datasets replaced by the \textit{BAOFULL} dataset, which comprises BAO measurements from the BOSS data release 11 (both CMASS and LOWZ samples), the 6dFGS survey, and the WiggleZ survey (see Tab.~\ref{tab:bao}). The relative constraining power of the geometric technique versus the shape approach  can be inferred by comparing the results of the first, second, third, fourth and fifth row to those shown in the third, fourth, sixth, eighth and ninth rows of Tab.~\ref{tab:tabmnutt}, respectively. The result is that, given our current analyses methods, geometrical information is more powerful than the shape one, see also the main text and Fig.~\ref{fig:shape_vs_geometry_planck}.}
\label{tab:tabmnubao}
\end{table*}

\begin{table*}[h!]
\begin{tabular}{|c?c|}
\hline
Dataset	& $M_\nu$ (95\%~C.L.)\\

\hline\hline
\textit{basepolBAO} $\equiv$ \textit{PlanckTT}+\textit{lowP}+\textit{highP}+\textit{BAOFULL} & $< 0.153$ eV \\	
\textit{basepolBAO}+$\tau0p055$ & $<0.118$ eV \\
\textit{basepolBAO}+$H073p02$ & $< 0.113$ eV \\
\textit{basepolBAO}+$H073p02$+$\tau0p055$ & $< 0.094$ eV \\
\textit{basepolBAO}+$H073p02$+$\tau0p055$+\textit{SZ} & $< 0.093$ eV \\
\hline
\end{tabular}
\caption{As Tab.~\ref{tab:tabmnubao}, but with the addition of \textit{highP}, referring to the small-scale CMB polarization anisotropies data. The relative constraining power of the geometric technique versus the shape approach can be inferred by comparing the results of the first, second, third, fourth and fifth row to those shown in the third, fourth, sixth, eighth and ninth rows of Tab.~\ref{tab:tabmnupol}, respectively. The result is that, given our current analyses methods, geometrical information is more powerful than the shape one, see also the main text and Fig.~\ref{fig:shape_vs_geometry_planckpol}.}
\label{tab:tabmnupolbao}
\end{table*}

\begin{table*}[h!]
\begin{tabular}{|c||c|c|c|}
\hline
Dataset&$M_\nu$ (95\%~C.L., \textit{3deg})&${\rm CL}_{\rm IH}$&$p_N/p_I$\\

\hline\hline
\textit{basepolPK}+$H073p02$+$\tau0p055$ & $< 0.109$ eV & $74 \%$ & $2.8:1$\\	
\textit{basepolPK}+$H073p02$+$\tau0p055$+\textit{SZ} & $<0.117$ eV & $71 \%$ & $2.4:1$\\
\textit{baseBAO}+$H073p02$+$\tau0p055$ & $< 0.115$ eV & $72 \%$ & $2.6:1$\\
\textit{baseBAO}+$H073p02$+$\tau0p055$+\textit{SZ} & $< 0.114$ eV & $72 \%$ & $2.6:1$\\
\textit{basepolBAO}+$\tau0p055$ & $< 0.118$ eV & $71 \%$ & $2.4:1$\\
\textit{basepolBAO}+$H073p02$ & $< 0.113$ eV & $72 \%$ & $2.6:1$\\
\textit{basepolBAO}+$H073p02$+$\tau0p055$ & $< 0.094$ eV & $77 \%$ & $3.3:1$\\
\textit{basepolBAO}+$H073p02$+$\tau0p055$+\textit{SZ} & $< 0.093$ eV & $77 \%$ & $3.3:1$\\
\hline
\end{tabular}
\caption{Exclusion C.L.s of the Inverted Hierarchy from our most constraining dataset combinations, obtained through a rigorous model comparison analysis. Only dataset combinations which disfavor the IH at $>70\%$~C.L. are reported. The first column lists the combination of cosmological datasets adopted, see Tab.~\ref{tab:data} for definitions. The second column reports the $95\%$~C.L. upper limit on $M_\nu$, obtained assuming the \textit{3deg} spectrum of three massive degenerate neutrinos. The third column reports ${\rm CL}_{\rm IH}$, the C.L. at which the IH is disfavored, calculated via Eq.~(\ref{clih}). Finally, the last column shows the relative posterior odds for NH versus IH, with the posterior probabilities for both mass orderings obtained via Eq.~(\ref{evidence}).}
\label{tab:inverted}
\end{table*}

\begin{acknowledgments}
SV, EG, and MG acknowledge Hector Gil-Mar\'{i}n for very useful discussions. SV and OM thank Antonio Cuesta for valuable correspondence. We are very grateful to the anonymous referee for a detailed and constructive report which enormously helped to improve the quality of our paper. This work is based on observations obtained with Planck (\href{http://www.esa.int/Planck}{www.esa.int/Planck}), an ESA science mission with instruments and contributions directly funded by ESA Member States, NASA, and Canada. We acknowledge use of the Planck Legacy Archive. We also acknowledge the use of computing facilities at NERSC. K.F. acknowledges support from DoE grant DE-SC0007859 at the University of Michigan as well as support from the Michigan Center for Theoretical Physics. M.G., S.V. and K.F. acknowledge support by the Vetenskapsr\aa det (Swedish Research Council) through contract No. 638-2013-8993 and the Oskar Klein Centre for Cosmoparticle Physics. M.L. acknowledges support from ASI through ASI/INAF Agreement 2014-024-R.1 for the Planck LFI Activity of Phase E2. O.M. is supported by PROMETEO II/2014/050, by the Spanish Grant FPA2014--57816-P of the MINECO, by the MINECO Grant SEV-2014-0398 and by the European Union’s Horizon 2020 research and innovation programme under the Marie Sk\l odowska-Curie grant agreements 690575 and 674896. O.M. would like to thank the Fermilab  Theoretical Physics Department for its hospitality. E.G. is supported by NSF grant AST1412966. S.H. acknowledges support by NASA-EUCLID11-0004, NSF AST1517593 and NSF AST1412966.
\end{acknowledgments}

%

\end{document}